\documentclass[iop]{emulateapj}

\usepackage[backref,breaklinks,colorlinks,citecolor=blue]{hyperref}
\usepackage[all]{hypcap}

\usepackage{graphicx}
\usepackage{amsmath}
\usepackage{textcase}

\newcommand \cv{\mathcal{V}}
\newcommand \Jinc{\mathrm{Jinc}}

\newcommand \kx{k_x}
\newcommand \ky{k_y}
\newcommand \bk{\bold{k}}

\newcommand \Rxy{$\mathfrak{R}(x,y)$}

\newcommand \eg {{\it e.g., }}

\newcommand \ie{{\it i.e.,}}
\newcommand \viz{{\it viz.,}}

\begin{document}
\accepted{November 8, 2014}
\shorttitle{Image plane NRM data analysis}
\shortauthors{Greenbaum et al.}
\title{An image-plane algorithm for JWST's non-redundant aperture mask data}
\author{ Alexandra Z. Greenbaum,}
	\affil{ Johns Hopkins University Department of Physics and Astronomy \\
	3400 N. Charles, Baltimore, MD 21218}
\author{ Laurent Pueyo,}
\author{ Anand Sivaramakrishnan\altaffilmark{1,2},}
	\affil{Space Telescope Science Institute, 3700 San Martin Drive, Baltimore, MD 21218}
\and
\author{Sylvestre Lacour}
	\affil{LESIA, CNRS/UMR-8109, Observatoire de Paris, UPMC, Universit\'e Paris Diderot \\ 5 place Jules Janssen, 92195 Meudon, France}
\altaffiltext{1}{Astrophysics Department, American Museum of Natural History,
                 79th Street and Central Park West, New York, NY 10024}
\altaffiltext{2}{Department of Physics and Astronomy, Stony Brook University, Stony Brook, NY 11794}

\begin{abstract}
The high angular resolution technique of non-redundant masking (NRM) or aperture
masking interferometry (AMI) has yielded images of faint protoplanetary
companions of nearby stars from the ground.  AMI on James Webb Space Telescope
(JWST)'s Near Infrared Imager and Slitless Spectrograph (NIRISS) has a lower
thermal background than ground-based facilites and does not suffer from
atmospheric instability.
NIRISS AMI images are likely to have $90$ - $95$\% Strehl ratio between 2.77
and 4.8~$\micron$.
In this paper we quantify factors that limit the raw point source contrast of
JWST NRM.
We develop an analytic model of the NRM point spread function which includes
different optical path delays (pistons) between mask holes and fit the model
parameters with image plane data.
It enables a straightforward way to exclude bad pixels, is suited to limited
fields of view, and can incorporate effects such as intra-pixel sensitivity
variations.
We simulate various sources of noise to estimate their effect on the standard
deviation of closure phase, $\sigma_{CP}$ (a proxy for binary point source
contrast).  If $\sigma_{CP} < 10^{-4}$ radians --- a contrast ratio of 10
magnitudes --- young accreting gas giant planets (\eg in the nearby
Taurus star-forming region) could be imaged with JWST NIRISS.
We show the feasibility of using NIRISS' NRM with the sub-Nyquist sampled
F277W, which would enable some exoplanet chemistry characterization.
In the presence of small piston errors, the dominant sources of closure phase
error (depending on pixel sampling, and filter bandwidth) are flat field errors
and unmodeled variations in intra-pixel sensitivity.  The in-flight stability
of NIRISS will determine how well these errors can be calibrated by observing a
point source.  Our results help develop efficient observing strategies for
space-based NRM.
\end{abstract} 

\section{Introduction}

Recent direct detections of exoplanets open a spectroscopic window into the
atmosphere and physics of young and adolescent exoplanets. They are an
important component for piecing together a complete picture of exoplanetary
formation and evolution,  and are complementary to indirect detections methods.
Young and nearby stars have already been surveyed from a few Astronomical Units
of physical separation outwards with direct imaging and coronagraphs on eight
meter class telescopes \citep{2013ApJ...773..179W, 2013ApJ...776....4N,
2013ApJ...777..160B, 2012AA...544A...9V} and are being surveyed at even higher
contrast with current \citep{2012SPIE.8447E..20O, 2012SPIE.8446E..1UM,
2010lyot.confE..44B, 2009SPIE.7440E..0OM} instrument surveys utilizing extreme adaptive optics
(ExAO). However, the close environs of young systems in stellar formation
regions are only accessible to ExAO systems using interferometric techniques such as
non-redundant mask (NRM) interferometry \citep{2011AA...532A..72L,
2012ApJ...745....5K, 2013ApJ...762L..12C, 2011AA...528L...7H}.  NRM imaging is
fundamentally limited by photon noise, so it yields moderate contrast.  By
comparison, coronagraphs (which suppress light from the bright central object)
are capable of delivering higher contrast than NRM, but their search area does
not reach as close in as that of NRM.  The two techniques are complementary.

NRM was first used to improve the angular resolution of filled-aperture
telescopes \citep{1986Natur.320..595B, 1987Natur.328..694H,
2000PASP..112..555T}.  Its improved dynamic range helped to probe the physics
of binaries at moderate contrast ratios \citep{2006ApJ...650L.131L,
2008ApJ...678..463I, 2010ApJ...715..724B, 2009ApJ...695.1183M,
2007ApJ...661..496M}.  More recently, NRM observations of star forming regions
have discovered structures associated with the birth of exoplanets
\citep{2012ApJ...745....5K,2013ApJ...762L..12C,2011AA...528L...7H}.  Routine
ground contrast ratio limits for NRM are $10^2 - 10^3$ with the deepest
contrast being $\Delta L'$=7.99 \citep{2011ApJ...730L..21H}.  Today instruments
combine NRM with ExAO systems
\citep{2010SPIE.7735E.266S,2011PhDT........54Z,2012SPIE.8445E..2GZ}.  These
facilities, together with wide bandpass polarization or integral field unit
spectroscopy (IFS) in the YJHK bands on the 8~m Gemini South telescope
\citep{Macintosh2014PNAS} as well as 2.5--5~\micron~NRM on the 40K James Webb
Space Telescope's Near Infrared Imager and Slitless Spectrograph (JWST NIRISS)
\citep{2012SPIE.8442E..2RD,2010SPIE.7735E.266S,2012SPIE.8442E..2SS,2013SPIE.88641L.56},
promise to extend planet formation science by enabling deeper dust penetration
at longer wavelengths. These new systems will enable detection of very young
(e.g. Taurus-Auriga star forming region), possibly accreting gas giant planets
at small separations accessible to NIRISS NRM \citep{2010PASP..122..162B}.

In spite of the wealth of recent observational results from NRM, the literature
does not include extensive discussion of the fundamental and practical limits
associated with the technique.  \cite{2011AA...532A..72L} discussed empirical
sensitivity limits of VLT NACO Sparse Aperture Masking (SAM), based on
experiments with the image plane fitting routine that we study here.
\cite{2010ApJ...724..464M} showed how NRM can be generalized to full aperture
imaging in the high Strehl ratio regime. \cite{2013MNRAS.433.1718I} discussed
some of the limiting factors of high contrast NRM observations, and
\cite{2011ApJ...730L..21H} conducted deep NRM observations of the known
multiple planetary system HR 8799. 
 
The purpose of this paper is two-fold. First, we continue the development of the
image plane approach to analyzing NRM data.  We address field of view, pixel
sampling, plate scale and pupil magnification stability, and some detector
properties.  We show that this method typically confirms the photon noise and
flat field accuracy limits presented by \cite{2013MNRAS.433.1718I}.  In
addition, we study other factors that limit NRM contrast --- requirements on
the spectral type match between a target and its calibrator star, and the
effect a finite spectral bandpass has on closure phase errors.
Second, we highlight factors that limit JWST NIRISS NRM, which fields a 7-hole
NRM \citep{2012SPIE.8442E..2SS}.  NIRISS has the best pupil image quality of
all the JWST instruments \citep{2008SPIE.7010E..98B}, which makes it JWST's
best-suited instrument for aperture masking interferometry.  In addition,
NIRISS's homogenous aluminum bench and optics should help achieve uniform
contraction of mechanical and optical surfaces as the instrument cools to its
operating temperature of about 40K.  NIRISS's all-reflective design philosophy
also mitigates against chromatic effects, which can be exacerbated by cryogenic
conditions.
Finally, some relevant properties of NIRISS NRM are described in the Appendix.

\section{Background}

\begin{figure} \centering
\includegraphics[scale=0.35]{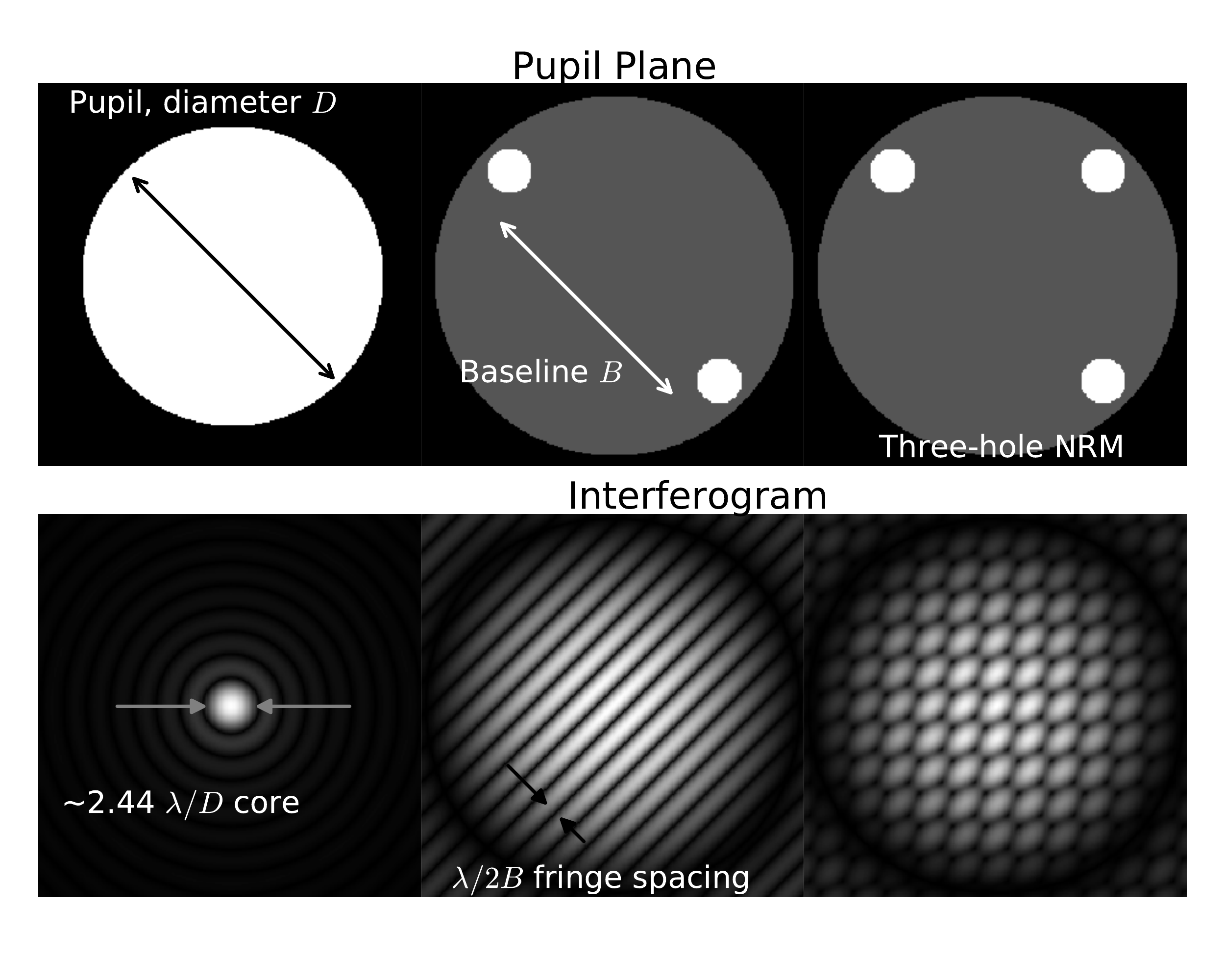}
\caption{\small \textbf{Pupil masks and their interferograms}.  Small holes
produce a large PSF envelope, fringed by interference through multiple holes.
The longest baselines provides finer resolution than a full aperture.  The
three hole pupil at right can provide a \textit{closure phase} measurement of a
celestial object.} \label{fig:interferogram} \end{figure}

(\autoref{fig:NRM}) A non-redundant mask is a pupil plane mask typically
located at a re-imaged pupil plane.  It possesses several usually identical
holes arranged so that no two hole-to-hole vectors are repeated (thus providing
a non-redundant set of \textit{baselines} in the pupil).  If its holes are
circular, with diameter $d$ when projected back to the primary mirror, at a
wavelength $\lambda$ its point-spread function (PSF) or \textit{interferogram}
is contained in an Airy pattern envelope with a first dark ring of diameter
$2.44 \ \lambda/d$ (\autoref{fig:interferogram}).
This envelope is modulated by fringes with half period $\theta =\lambda/2B$ for
each baseline.  Here $B$ is the hole separation.
\autoref{fig:NRM} shows the JWST NIRISS mask with seven hexagonal holes,
and its PSF.

\begin{figure} \centering{
\includegraphics[height=1.5truein]{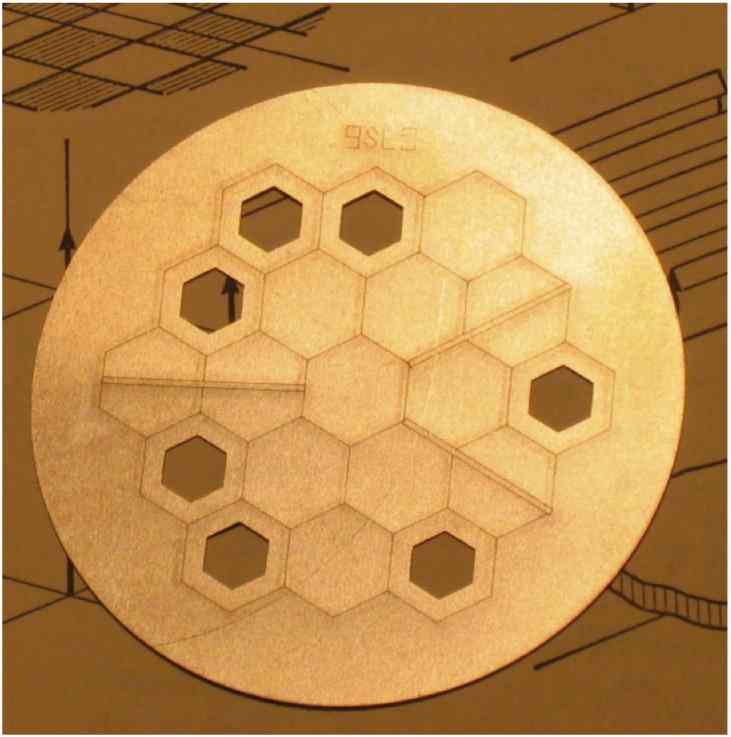}
\includegraphics[height=1.5truein]{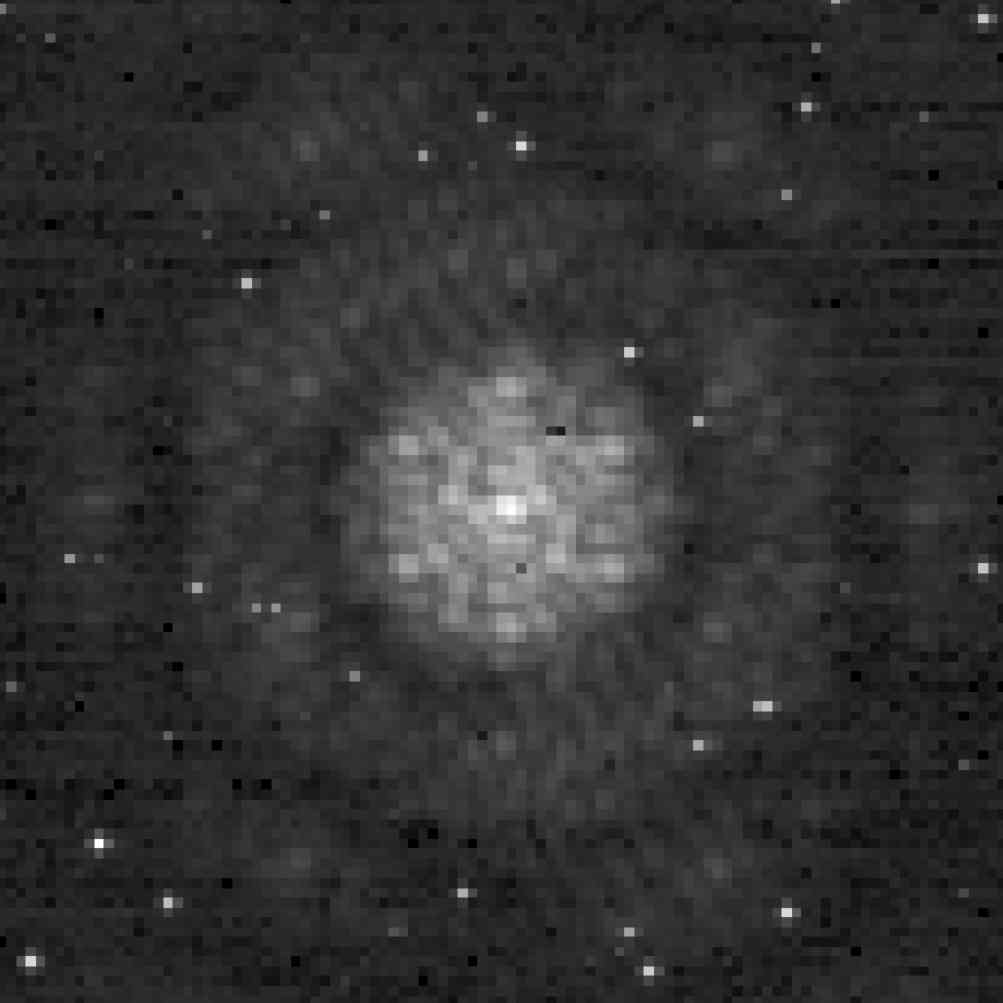}} \caption{\small
Non-redundant mask for JWST's NIRISS pupil wheel \citep{2010SPIE.7731E.126S}
and its PSF (or interferogram) with NIRISS F430M from cryogenic vacuum tests in
November 2013 \citep{AlexSPIE2014}. The interferogram's fine structure is due
to the 21 baselines generated by the 7 holes. The PSF envelope reflects the
hexagonal hole shape.} \label{fig:NRM}
\end{figure}

The Fourier transform of the detected in-focus two-dimentional image intensity
array is the array of \textit{complex visibility}, $\cv$.  Because of the
baselines' non-redundancy, the fringe amplitude and phase for each baseline or
``two hole interferometer'' component in the NRM can be measured unambiguously.
The array of complex visibilities for a point source through unaberrated optics
is the autocorrelation of the pupil mask. The resulting array of complex
visibilities form localized \textit{splodges} \citep{2006ApJ...650L.131L} of
signal in the transform domain --- conceptually one independent splodge (or a
spodge and its dependent, Hermitian ``mirror splodge'') per baseline.
Numerical Fourier data analysis approaches measure fringe phases and fringe
amplitudes, often at the peak of each splodge amplitude
\citep{2000PASP..112..555T, 2006ApJ...650L.131L}.  When using a finite
bandwidth filter, selecting a single amplitude and phase to characterize a
polychromatic fringe implicitly averages over the bandpass.  Furthermore,
since windowing in the image plane leads to convolution in the Fourier domain,
this induces a second form of averaging within a splodge.  Our image plane
approach avoids this second form of averaging, but it does perform a
conceptually similar averaging over the bandpass.  In the absence of wavefront
aberration, fringe phases for an on-axis point source are zero.  Information on
source structure is contained in the fringes that are extracted from the image.

%%%
The non-redundancy of baselines in the pupil leads to constraints on the complex
fringe visibilities.
A \textit{closure phase} (the cyclic sum of fringe phases around the baselines
formed by three holes (top right, \autoref{fig:interferogram})) is
insensitive to constant wavefront delays (pistons) over the holes.  The fringe
formed by interference of holes $i$ and $j$ has a fringe phase $\phi_{i,j}$
which is proportional to the wavefront delay between holes
$\phi_{i,j} \equiv \phi_{j} - \phi_{i}$.  For a point source (in the absence of
higher order aberrations) \citep[e.g.][]{1988AJ.....95.1278R}:
\begin{eqnarray} 
\Delta\phi_{1,2}+\Delta\phi_{2,3}+\Delta\phi_{3,1} &=& \nonumber \\
(\phi_{1}-\phi_{2})+(\phi_{2}-\phi_3)+(\phi_{3}-\phi_1) &=& 0
\label{eq:closurephase} 
\end{eqnarray}
Full-aperture images do not yield closure phases, but sufficiently high Strehl
ratio images possess certain constrained linear combinations of phases of the
Fourier transform of the image \citep{2010ApJ...724..464M,
2011SPIE.8151E..33M}.  These combinations, or \textit{kernel phases}, are
useful for model fitting data when wavefront aberrations are below
$\sim 1$~radian \citep{2010ApJ...724..464M, 2013MNRAS.433.1718I, 2013ApJ...767..110P}.

An $N$-hole mask has $N(N-1)$/2 baselines,  $N(N-1)(N-2)/6$ closure phases, and
$(N-1)(N-2)/2$ independent closure phases.  Empirically, achievable dynamic range is
approximately the inverse of the standard deviation of closure phase error,
$1/\sigma_{CP}$ \citep{2011AA...532A..72L}. 

Closure phases of centro-symmetric sky brightness distributions are zero.
Binary or multiple point source models are fit to closure phase data to provide
information on structure as fine as $\lambda/2B$.  Instrumental contributions
to closure phases are measured (in principle) by observing a point source.
These contributions are then subtracted from a target's closure phases.
Instrument stability between target and calibrator leads to improved NRM
performance.  In addition to fringe phases, a space telescope is likely to
provide stable fringe amplitudes.  Closure amplitudes (a ratio of amplitudes of
fringes formed by four holes \citep{1986isra.book.....T}) are useful in simple
model fitting using space-based NRM data, thereby extending NRM model fitting
to include centro-symmetric structure such as circular disks.  However,
\citet{2014ApJ...783...73F} use simulated noisy NIRISS NRM data to extract the
fringe amplitudes and phases which they then use to recreate the input target
scene with interferometric resolution.  They found that enforcing closure
quantities on image plane data leads to an increase in spurious image artifacts.

Currently numerical Fourier methods are the most common approach to NRM data
analysis \citep[e.g.][]{2003RPPh...66..789M,2000PASP..112..555T,
2012ApJ...745....5K, 2008ApJ...678..463I}.  This is suited to fields of view
that encompass the first few Airy rings of the NRM PSF's ``primary beam'' (the
diffraction pattern of a single hole), and pixel scales that are significantly
finer than $\lambda/2D$.  Palomar Hale's PHARO, Keck-NIRC2, and VLT's NACO all
possess 3-5 pixels per resolution element
\citep{2004ApJ...617.1330M,2011ApJ...726..113I,2000PASP..112..555T,2011ApJ...726..113I,NACO}.
With such super-Nyquist fine pixellation, Fourier methods easily identify and
interpolate over isolated bad pixels \citep{2013MNRAS.433.1718I}.

Diffraction-limited exoplanet imagers deploying state-of-the-art ExAO systems
now feed IFSs \citep{2012SPIE.8447E..20O,Macintosh2014PNAS,
2008SPIE.7014E..41B}.  These imaging spectrographs typically have limited
fields of view since several detector pixels are required for each image plane
pixel spectrum, and the angular extent of each image plane lenslet is at or
below the diffraction limit of the telescope, so the instruments are often
limited by the number of available detector pixels.  NRM on these hyperspectral
imagers --- Palomar's P1640 \citep{2011PhDT........54Z, 2012SPIE.8445E..2GZ}
and Gemini Planet Imager \citep{2010SPIE.7735E.266S, 2013SPIE.88641V.66} ---
must deal with this limitation.  An image plane based approach
\citep[e.g.][]{2011AA...532A..72L, Cheetham:12, 2013SPIE.88641L.56} is
insensitive to these restrictions on the field of view.

Future space-based NRM on JWST NIRISS \citep[e.g.][]{2009SPIE.7440E..30S,
2009astro2010T..40S, 2010SPIE.7731E.126S, 2012SPIE.8442E..2SS,
2013SPIE.88641L.56} is implemented on coarse pixel scales.
Under these conditions a numerical Fourier data reduction approach may require
more data in order to reduce contamination by bad pixels.  This is more
relevant to coarse --- barely or sub-Nyquist --- pixel scales.  Dithering to
fill the image plane pixels with valid data decreases observing efficiency and
complicates estimates of noise.  An image plane based approach sidesteps the
requirement of knowing every pixel value in the image.  The image plane
approach is also robust to detector non-linearities that may occur at the
centers of NRM images, since suspect pixel data can be discarded.
JWST NIRISS's coarse pixel scales also increase its sensitivity to non-uniform
sensitivity within a pixel (intra-pixel sensitivity, or IPS), and
pixel-to-pixel variations in IPS \citep{2008SPIE.7021E..70H}.  Image plane data
reduction can take IPS into account, with a map of measured variations or a
model of the pixel sensitivities \citep{2013SPIE.88641L.56}.

\section{Image Plane Modeling} 
We assume the image plane complex amplitude induced by a point source at infinity
is described by the Fourier transform of the aperture transmission function
(\ie\ the Fraunhofer approximation).  If functions F and f
are a Fourier transform pair, we write $F \overset{F.T.}\rightleftharpoons f$.
We develop a polychromatic image plane model tailored to JWST NIRISS's seven
hole NRM (\autoref{fig:NRM}).  Each hole is a hexagon, which, when projected
to the JWST primary mirror, has a flat-to-flat distance of approximately 0.8~m.
Our model can be adapted to arbitrary hole locations and polygonal hole shapes
\citep[e.g.][]{2013SPIE.88641V.66}.  Here we treat circular
holes with diameter $d$ or hexagonal holes with flat-to-flat distance $D$
(\autoref{fig:hex}),
utilizing a closed form for the Fourier transform of a hexagon
\citep{2005ApOpt..44.1360S}, while noting that other more specialized
derivations for this exist in the literature \citep{2003ApOpt..42.3745T}.  
We extend the work of \citet{2005ApOpt..44.1360S} to include limiting values for
the analytical expression's three singular lines and singular central point.

We calculate the monochromatic NRM PSF at a wavelength $\lambda$ analytically,
and construct polychromatic PSFs by summing appropriately weighted monochromatic PSFs
on a finely sampled numerical grid.  We then bin this finely sampled image to the detector
pixel scale to simulate a pixelated noiseless NRM PSF.

We denote the pupil transmission function by $A(\bold{x})$.
A hole with a transmission function $A_h(\bold{x})$ produces an image plane complex
amplitude $a_h(\bold{k})$ (where $ a_h \overset{F.T.}\rightleftharpoons A_h$)
and a PSF $P = a_h\,a_h^*$
(where $^*$ indicates complex conjugation). 
If the pupil plane vector $\bold{x} = (x,y)$ is in units of the wavelength of the monochromatic light,
the image plane (or spatial frequency) vector $\bold{k} = (k_x,k_y)$ is in cycles across the pupil. 
$P(\bold{k})$ is the \textit{primary beam}, by analogy with radio interferometry,
and is the envelope of the NRM PSF.
Vector baselines create the finer scale fringing in the NRM PSF.

\subsection{Circular mask holes} 
A circular aperture's transmission function is
  \begin{eqnarray}
   ^{2}\prod{(\bold{x})} &=& \left\{
     \begin{array}{lr}
       1 &, \ \ \   r <\frac{1}{2}\\
       0 &, \ \ \   r \geq \frac{1}{2}
     \end{array}
   \right. 
   \label{eq:tophat}
\end{eqnarray}
where  $r = \sqrt{x^{2} + y^{2}}$.
The transmission function of a mask with $N$ identical circular holes
centered at $\{\bold{x}_{i},\  i=1, ..., N\}$ is
  \begin{eqnarray}  
\label{eq:aperture} A(\bold{x}) &=& \sum_{i=1}^N \
^{2}\prod(\frac{\bold{x}-\bold{x}_{i}}{d_\lambda})
\end{eqnarray}
(where $d_\lambda \equiv d/\lambda$).
The image plane complex amplitude of an on-axis monochromatic point source
observed through this mask is
  \begin{eqnarray}  \label{eq:imagefield}
       a(\bold{k})  &\overset{F.T.}\rightleftharpoons&  A(\bold{x}).
  \end{eqnarray}
Following the nomenclature of phase retrieval work on the Hubble Space Telescope,
we call $a(\bold{k})$ the \textit{amplitude spread function} (ASF).

Invoking Fourier shift and scaling theorems,
\begin{eqnarray} \label{eq:jinc}
^{2}\prod(\frac{\bold{x}-\bold{x}_{0}}{d_\lambda})
\overset{F.T.}\rightleftharpoons (d_\lambda)^{2} \Jinc(k d_\lambda)e^{- i
\bold{k} \cdot \bold{x}_{0}}
\end{eqnarray}
where $\Jinc(k) \equiv J_{1}(\pi k)/2 k$  is the transform of the circular
transmission function.  Here $J_1$ is the Bessel function of the first kind, of order 1.
The phase gradient term $e^{- i \bold{k} \cdot \bold{x}_{0}}$ reflects
the shift of the hole's origin to $\bold{x}_0$.

\subsection{Hexagonal mask holes}

We denote the hexagonal hole Fourier transform by $a_{\mathrm{hex}}(\bk)$.
Following \cite{2005ApOpt..44.1360S},
$g(\kx,\ky)$ is the Fourier transform of one half
of a hexagon that is bisected from one corner to it diametrically opposite
corner  (\autoref{fig:hex}):
\begin{eqnarray} \label{eq:hextransform}
g(\kx, \ky) &=& \frac{\exp[-i\pi D (\frac{2\kx}{\sqrt{3}}+\ky)]}{4\pi^2(\kx^3 -
3\kx\ky^3)} \nonumber (\sqrt{3}\kx-3\ky ) \\ 
&\times& (\{\exp(i\pi D\sqrt{3}\kx) - \exp[i\pi D(\frac{4}{\sqrt{3}}\kx +
\ky)]\} \nonumber \\
&+& (\sqrt{3}\kx + 3\ky)[\exp(i\pi D\kx/\sqrt{3}) - \exp(i\pi D\ky)]) \nonumber
\\
a_{\mathrm{hex}}(\ky, \kx) &=& g(\kx, \ky) + g(\kx, -\ky).
\end{eqnarray}
The function $g$ has numerical singularities along three lines,
$\kx = 0$ and $\kx = \pm \sqrt{3} \ky$.
The limiting behavior along $\kx=0$ and at the origin is:

\begin{figure}
\centering
\includegraphics[scale=0.4]{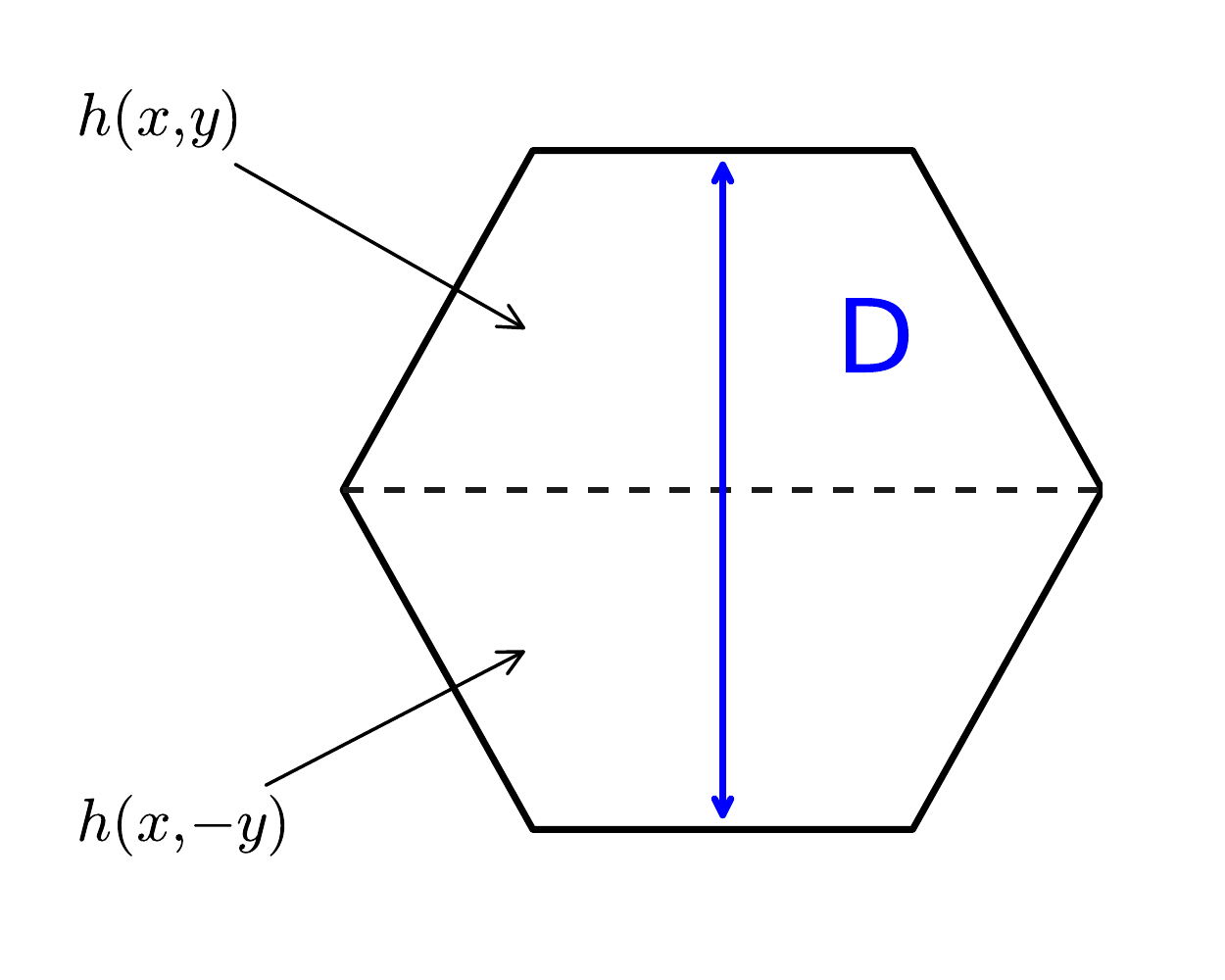}
\caption{\small In \cite{2005ApOpt..44.1360S} 
$D$ is the flat-to-flat distance. The hexagon is split into two symmetric
parts, $a_{\mathrm{hex}}(x,y)$ and $a_{\mathrm{hex}}(x,-y)$, whose transforms,
$g(k_x, k_y)$ and $g(k_x,-k_y)$ are computed analytically 
(\autoref{eq:hextransform}).} \label{fig:hex}
\end{figure}
\begin{eqnarray}
&g&(0,\ky) = \frac{e^{-i D\pi\ky}}{2\sqrt{3}\pi^2\ky^2} \times \nonumber \\
&&(-1+ i D \pi\ky + e^{i D \pi\ky} - 2i D \pi\ky e^{i D\pi\ky}) \\
&g&(0,0) = \frac{\sqrt{3}D^2}{4}.
\end{eqnarray}
Values along the other two lines can be found by invoking symmetry
arguments, and replacement with the appropriate limiting value taken from
the $k_x=0$ line.

\subsection{Interference between holes}

In the absence of wavefront error the ASF  of 
a mask with $N$ identical holes centered at $\{\bold{x}_{i}, i=1, ..., N\}$ is
\begin{equation}
\sum_{i=1}^N A(\bold{x} - \bold{x}_i) \overset{F.T.}\rightleftharpoons
a(\bold{k}) = a_h(\bold{k}) \sum_{i=1}^N e^{- i \bold{k} \cdot \bold{x}_{i}}
\label{eq:ASFperfect}
\end{equation}
($a_h(\bold{k})$ is a single hole ASF).
The mask's point spread function is
\begin{eqnarray} \label{eq:psf}
p(\bold{k}) = a(\bold{k})a^{*}(\bold{k})=P(\bold{k}) \sum_{i=1}^N \sum_{j=1}^N
e^{- i \bold{k} \cdot( \bold{x}_{i}-\bold{x}_{j})}
\end{eqnarray}
or
\begin{eqnarray} \nonumber
p(\bold{k}) = P(\bold{k}) \{N + e^{- i \bold{k} \cdot (\bold{x}_{1} -
\bold{x}_{2})} + e^{ i \bold{k} \cdot (\bold{x}_{1} - \bold{x}_{2})}. \\
\nonumber
+ e^{- i \bold{k} \cdot (\bold{x}_{1} - \bold{x}_{3})} +e^{ i \bold{k} \cdot
(\bold{x}_{1} 
- \bold{x}_{3})} + ...\}
\end{eqnarray}
(which is real and nonnegative for any $\bold{k}$).  The flux in this
image is the two-dimensional integral $\int N P(\bold{k}) d\bold{k}$, taken
over the entire $\bold{k}$ plane.
We rewrite the PSF as \\
\begin{equation} \label{eq:fringes}
p(\bold{k})=P(\bold{k})\{N + \sum_{i<j} 2\cos{( \bold{k} \cdot (\bold{x}_{i} -
\bold{x}_{j}))}\}
\end{equation}
which shows the separate roles the vector baselines and the primary beam 
play in the morphology of a point source's interferogram.

Wavefront errors $\{\phi_i,\  i=1,...,N\}$ that are constant within each of the
apertures decenter each fringe by $(\phi_i - \phi_j)$. Such errors are termed
\textit{pistons}.  Pistons do not move the image centroid, since the intensity
centroid is the mean of the phase gradient over the (uniformly illuminated)
pupil forming an in-focus image \citep{1982JOSA...72.1199T}, and piston errors
do not change the mean wavefront slope.  A piston difference between two holes
shifts the fringe away from the image centroid (or \textit{pointing center}) by
an angle, the \textit{fringe phase}.  A shift from one fringe maximum to the
next is interpreted as an angle of $2\pi$.  Given JWST NIRISS' anticipated
image quality during normal operations, we expect fringe phases of point source
NRM images to lie well inside the half-open interval $(-\pi,\pi]$.  This
removes any technical difficultes associated with a fringe phase wrapping
around $2\pi$.
We stress that fringe phases are not the argument of a `phasor' associated with
the complex amplitude of an electromagnetic wave.
The expression for the interferometric PSF in the presence of only piston
errors is

\begin{eqnarray}  \label{eq:PSF}
   p(\bold{k}) &=&P(\bold{k}) ~  \sum_{i=1}^N \sum_{j=1}^N  
   e^{- i \bold{k} \cdot( \bold{x}_{i}-\bold{x}_{j}) +  i(\phi_{i} - \phi_{j})}  \nonumber \\
    &=&P(\bold{k}) \Big\{N + \sum_{i<j} 2\cos{( \bold{k} \cdot (\bold{x}_{i} - \bold{x}_{j})
    +(\phi_{i}-\phi_{j}))}\Big\} \nonumber \\ \nonumber
    &=&P(\bold{k}) \Big\{N + \sum_{i<j}2\Big(\cos{( \bold{k} \cdot (\bold{x}_{i} 
    - \bold{x}_{j}))}\cos{(\phi_{i}-\phi_{j})}  \\ 
      &&-\sin{( \bold{k} \cdot (\bold{x}_{i} - \bold{x}_{j}))}\sin{(\phi_{i}-\phi_{j})}\Big)\Big\} .
\end{eqnarray}

\subsection{The JWST NRM PSF}
For JWST NIRISS's 7-hole hexagonal mask, (\autoref{eq:PSF}) gives
\begin{eqnarray} \label{eq:model} \nonumber
p(\bold{k}) = P(\bold{k}) ~ \big\{7+ 2\cos{(
\bold{k}\cdot(\bold{x}_{1}-\bold{x}_{2}))}\cos{(\Delta\phi_{1,2})} \\ \nonumber 
-2\sin{( \bold{k}\cdot(\bold{x}_{1}-\bold{x}_{2}))}\sin{(\Delta\phi_{1,2})}  \\
\nonumber
+ 2\cos{( \bold{k}\cdot(\bold{x}_{1}-\bold{x}_{3}))}\cos{(\Delta\phi_{1,3})} \\
\nonumber
-2\sin{( \bold{k}\cdot(\bold{x}_{1}-\bold{x}_{3}))}\sin{(\Delta\phi_{1,3})} \\
+...\big\} \label{eq:model}.
\end{eqnarray} 

With this closed form rapid calculation of monochromatic and polychromatic
PSFs on a fine scale is straightforward.

\subsection{Linear fit} \label{sec:fit} 
Piston differences enter into \autoref{eq:model} as coefficients of the
sines and cosines describing the baselines' fringes.  NIRISS's 7-hole mask has
42 such fringe coefficients--- $\cos{\Delta\phi_{i,j}}$'s and
$\sin{\Delta\phi_{i,j}}$'s, which we rename $a_{ij}$'s and $b_{ij}$'s,
respectively.  Two additional parameters are required to match the model to
data: the average flux per hole, $F$, and a DC offset $C$:
\begin{eqnarray} \label{eq:complete} 
&& F ~ P(\bold{k})~\{N + \nonumber \\
&\sum_{i<j}& \ 2[\cos{( \bold{k} \cdot (\bold{x}_{i} -
\bold{x}_{j}))}\cos(\Delta\phi_{i,j}) \nonumber \\ 
&& \ - \sin{( \bold{k} \cdot (\bold{x}_{i} -
\bold{x}_{j}))}\sin(\Delta\phi_{i,j})]\}+ C. \end{eqnarray}
These 44 parameters can be estimated from image plane pixel data by using an
unweighted \textit{linear} least squares minimization of the quantity 
\begin{equation} \label{eq:minimize} || \mathrm{data} - \mathrm{model}(a_{ij},
b_{ij}, F, C)||,  \nonumber 
\end{equation} 
and performing a matrix inversion to recover the parameters.  We did not detect
significant improvent of a noise-weighted fit over an equally weighted fit,  so
we use the latter.  The piston differences, or fringe phases, are found with
\begin{equation} \label{eq:tan} \Delta\phi_{ij} =
\mathrm{arctan}(b_{ij}/a_{ij}).  \end{equation}
For uniformly transmissive optics throughput, no scattered light, no
significant high spatial frequency wavefront errors, and perfect detectors we
expect the trigonometric identity
$$b_{ij}^2 +a_{ij}^2 = 1$$ to hold when imaging a point source.  Model
parameters derived from fitting real data rarely obey this identity.  Instead,
we obtain  the square of the $ij^{\mathrm{th}}$ fringe visibility:
\begin{equation} \label{eq:vissq} b_{ij}^2 +a_{ij}^2 = \cv_{ij} \cv^*_{ij}.
\end{equation}
Target structure further reduces fringe visibility.  We calculate fringe
visibilities in our simulated data sets by measuring coefficients
$\{a_{ij},b_{ij}\}$.  We calculate all 35 possible closure phases in NIRISS's
7-hole NRM.  Only 15 of these are independent measurements.

We evaluate our model PSF on a $3\times3$ sub-pixel grid
(unless otherwise noted) so we can study sub-pixel effects,
and then bin the array to the detector pixel scale.
A full pupil
distortion model was not used in this study, although real data will require
detailed knowledge of the NRM-to-primary mapping.

A polychromatic model is generated with an appropriately weighted sum of each
monochromatic fringe model, given the bandpass profile.  
\begin{eqnarray} \nonumber
	model = \sum_{\lambda}
	F_\lambda ~ P(\bk_\lambda)~\{N + && \\
	\sum_{i<j}2[\cos(\bk_{\lambda}\cdot(\bold{x}_{i} -
\bold{x}_{j}))\cos(\Delta\phi_{i,j}) &&
\nonumber\\ 
	- \sin(\bk_{\lambda}\cdot(\bold{x}_{i} -
	  \bold{x}_{j}))\sin(\Delta\phi_{i,j})]\}. &&
\label{eqn:polymodel} \end{eqnarray} 
In the presence of non-zero piston error the model in
\autoref{eq:complete}  does not fit polychromatic data perfectly, because
piston error scales inversely with wavelength.  This means that the fringes'
coefficients,
	  $ \cos(\Delta\phi_{i,j})$ and
	  $ \sin(\Delta\phi_{i,j}) $,
themselves vary with wavelength, but our fit keeps these coefficients
constant over the bandpass.  The narrower the fractional bandwidth of the
filter, the smaller the variation of these coefficients.  This problem is
common to both the image plane as well as the numerical Fourier approach to NRM
data analysis.  The least squares solution (\autoref{eq:tan} and
\autoref{eq:vissq}) produces an estimate of fringe phase and amplitude that
describes some average over the bandpass.  We use this estimate in our
polychromatic studies.

\subsection{Applicability of the model} \label{sec:applicability} 
NRM is suited to wavefronts that are smooth over each hole in the mask.  Our
model assumes flat wavefronts over each hole, which averages over fine
wavefront structure in some manner.  A Fourier approach windows image data,
which also averages phase and visibility information (since image plane
windowing is a convolution in the Fourier domain).  However, the two approaches
propagate image plane noise differently.  We discuss the effects of high
spatial frequency wavefront error in section \autoref{sec:WFE}.

\autoref{tab:NIRISSfilters} describes NIRISS NRM filters relevant to exoplanet 
studies. We study filter bandpass, source temperature, and spectral type effects 
using our polychromatic PSF model.
When using F277W and F380M, NIRISS' 65~mas square detector pixels are coarser 
than Nyquist-sampled.
%%%
\capstartfalse
\begin{deluxetable}{cccc}
%\tablewidth{2.9truein}
\tablewidth{0pt}
\tablecaption{JWST NIRISS Filters for NRM\label{tab:NIRISSfilters}}
\tablehead{\colhead{Filter} & \colhead{$\lambda_C/\micron$} &
           \colhead{$\delta\lambda/\lambda$} & \colhead{$\lambda/2D$  mas } 
}
\startdata
F480M &  4.8 & 0.08 & 76  \\
F430M &  4.3 & 0.05 & 68  \\
F380M &  3.8 & 0.05 & 60  \\
F277W &  2.7 & 0.25 & 44  \\
\enddata
\tablecomments{ $\lambda_C$ is the filter central wavelength, and $\delta \lambda$
	is the half-power width of the filter. The Nyquist pixel scale $\lambda/2D$ 
	uses the nominal equivalent area JWST mirror diameter $D=6.5$.  NIRISS's 
	pixel scale is 65~mas.}
\end{deluxetable}
\capstarttrue

\section{Photon noise, flat field error, and intra-pixel sensitivity}
\label{sec:data}

\begin{figure}
\centering
\includegraphics[scale=0.3, angle=90]{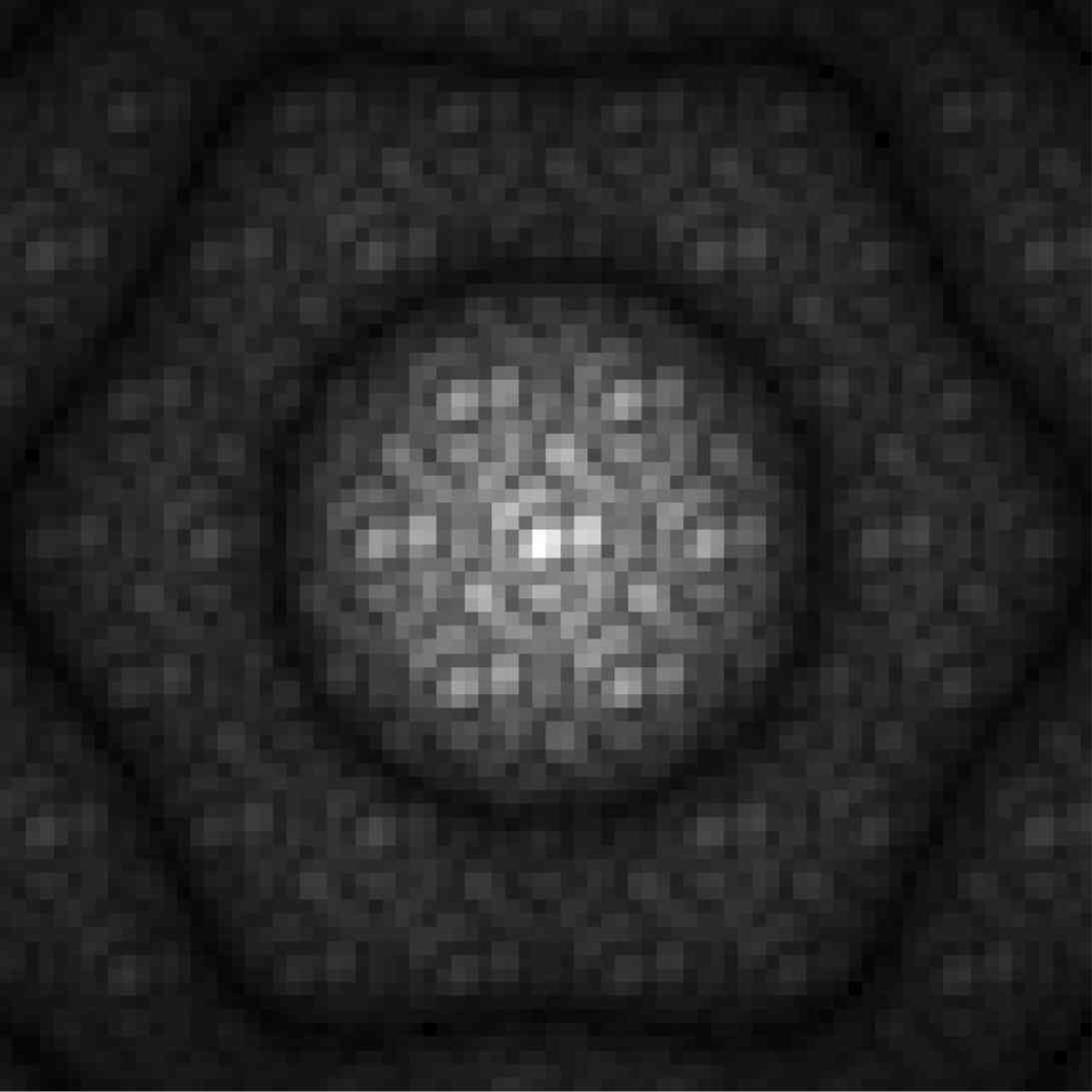}
\caption{\small Asserting the static pistons from \autoref{tab:simpistons} in this
simulated PSF produce asymmetric features.
}
\label{fig:pistons}
\end{figure}

We inserted piston wavefront errors over the holes
(\autoref{tab:simpistons} and \autoref{fig:pistons}) 
to examine their effects on simulated monochromatic images.
Our pistons are all smaller than $\lambda/4$, which avoids phase wrapping.

\capstartfalse
\begin{deluxetable}{c}
%\tablewidth{3.0truein}
\tablewidth{0pt}
\tablecaption{Simulation static pistons\label{tab:simpistons}}
\tablehead{\colhead{Piston in waves at 4.3~\micron  } }
\startdata
$+0.02884$ \\
$-0.06150$ \\ 
$+0.12400$ \\
$-0.02040$ \\ 
$+0.01660$ \\
$-0.03960$ \\
$-0.04780$ \\
\enddata
\tablecomments{Our simulations use a set of uniformly distributed, random 
static pistons with a mean of zero and standard deviation 0.06 waves.  Anticipated JWST NIRISS
rms wavefront error at 4.3~\micron\ is approximately $\lambda/30$.}
\end{deluxetable}
\capstarttrue

\begin{figure}
\centering
\includegraphics[scale=0.6]{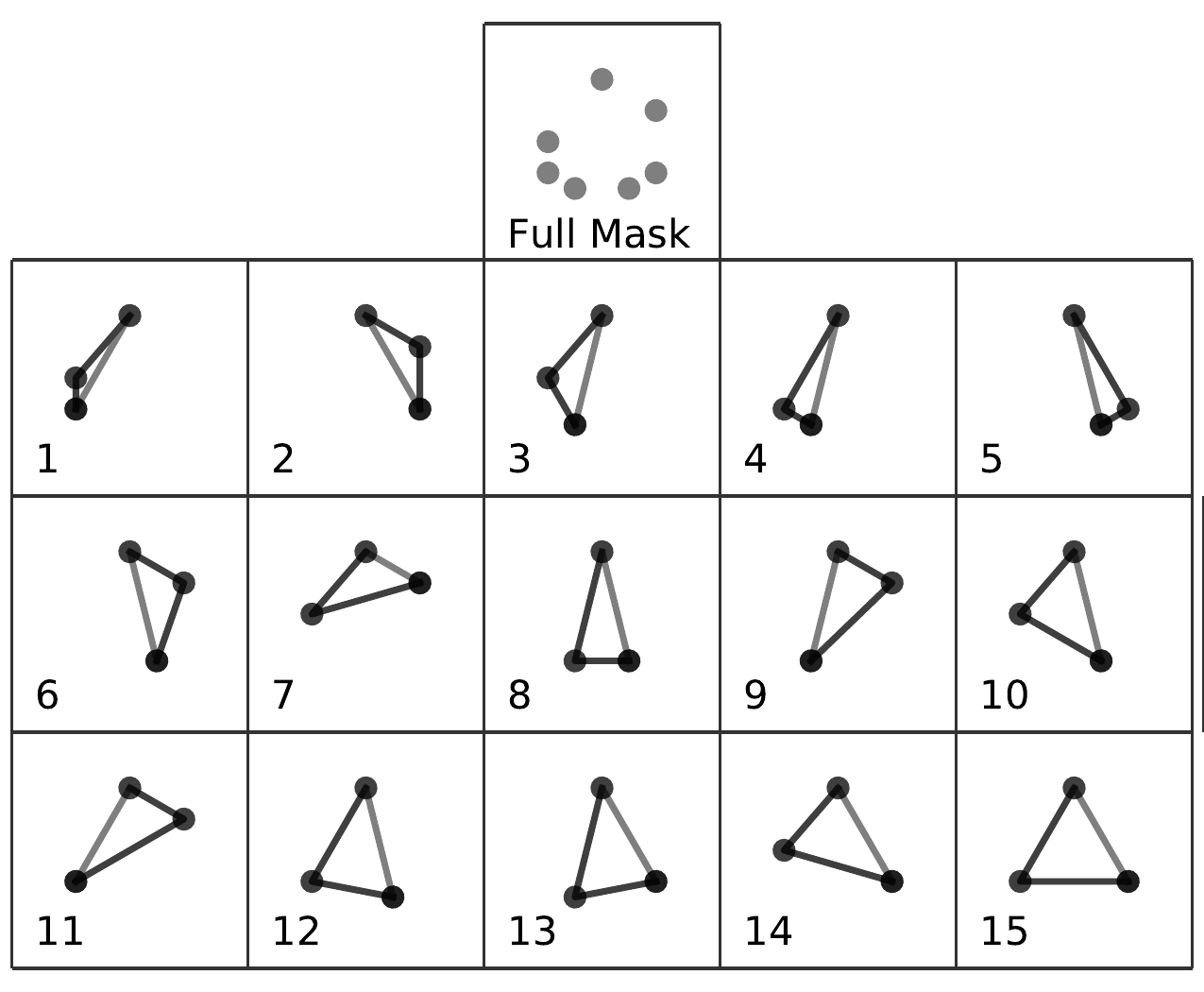} \caption{\small \textbf{Closure
triangles}: The set of 15 independent closure triangles corresponding to the
figures in \autoref{sec:data}, \autoref{sec:polychromatic}, and \autoref{sec:WFE},
ordered by increasing perimeter.} \label{fig:triangles}
\end{figure}

We generate monochromatic 4.3~\micron\ images either 3 or 5 times finer than
the NIRISS pixel scale prior to binning to its detector scale and performing a
least squares determination of the 44 model parameters.  We measure
closure phase standard deviation for different noise parameters for a set of 15
independent closure triangles (\autoref{fig:triangles}). In the absence of
added noise our measured closure phases were numerically indistinguishable from
zero.

We find that the quality of the fit does not change significantly with field of
view (i.e. the number of pixels used). We used data within
the first dark Airy ring of the primary beam in all simulations. 

We distinguish between two types of closure triangle response to different
sources of noise or error:
\begin{itemize} \item \textit{Baseline-independent} behavior limits contrast at
all spatial frequencies similarly. 
\item \textit{Baseline-dependent} behavior varies with baseline length and
therefore closure triangle. 
This behavior preferentially limits contrast at higher frequencies, or smaller
angular resolution.  \end{itemize}

\subsection{Photon noise}
\begin{figure} \centering \includegraphics[scale=0.3]{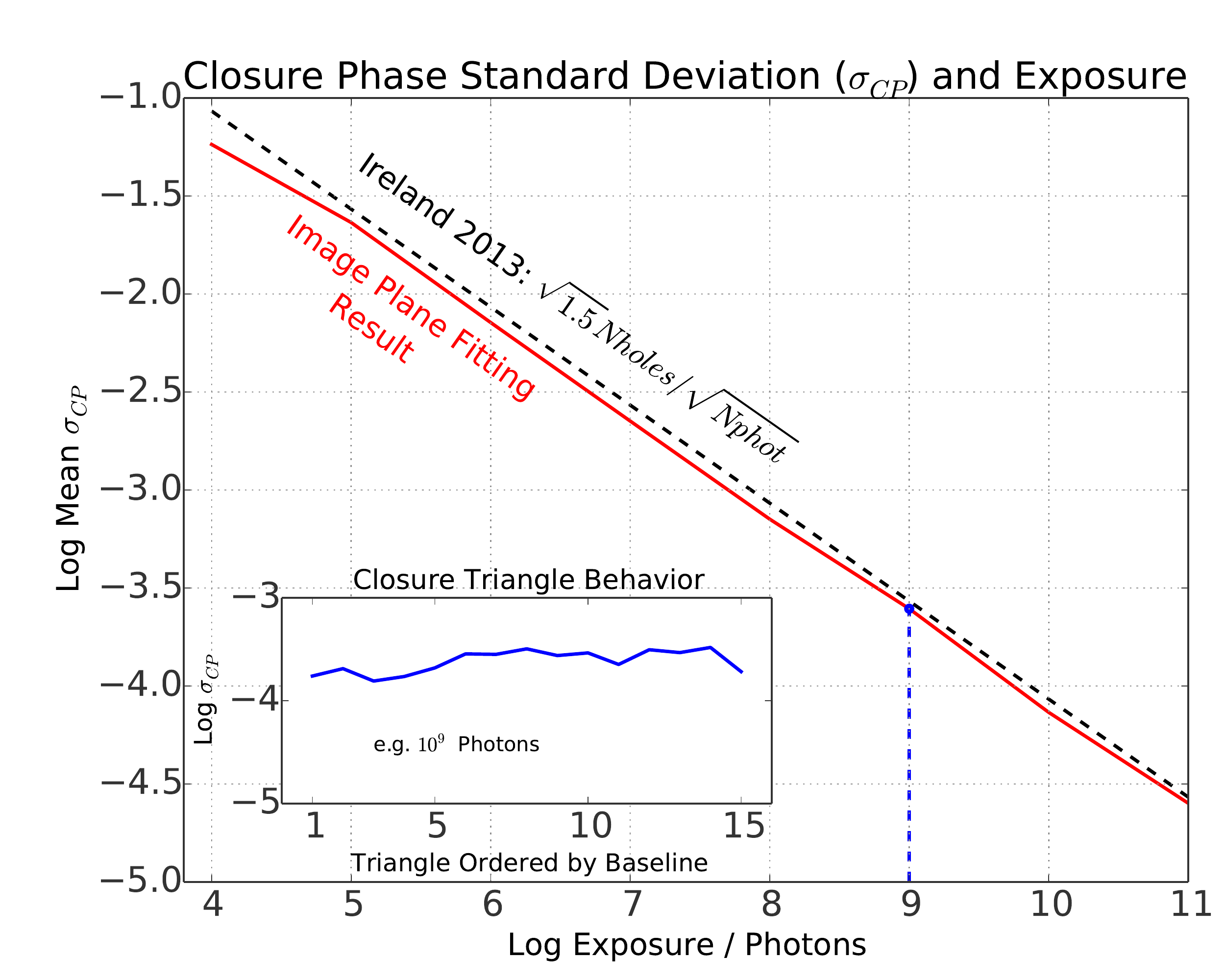}
\caption{\small \textbf{Exposure time (monochromatic):} The mean closure phase
standard deviation for a range of exposure.  The solid red line shows our
simulation results; the dotted black line displays the noise limit of
\cite{2013MNRAS.433.1718I}.  The inset shows one example of $\sigma_{CP}$ for
each of the 15 triangles in \autoref{fig:triangles} for an exposure with
$10^9$ photons.  Photon noise induces \textit{baseline-independent} errors.}
\label{fig:poisson} \end{figure}

We investigate a range of exposures, from $10^4$ to $10^{11}$ photoelectron
counts (assuming coadding of multiple exposures regardless of pixel well
depth).  We calculate the standard deviation of each of the 35 closure phases
over 25 independent realizations, and plot the mean of these standard
deviations, $\sigma_{CP}$, as the solid line in \autoref{fig:poisson}.  Our
results are consistent with the \cite{2013MNRAS.433.1718I} result $$\sigma_{CP}
= \sqrt{1.5}\frac{Nholes}{\sqrt{Nphot}}$$ indicated by the dotted line.  The
inset plot in \autoref{fig:poisson} displays the behavior of the closure
phases plotted in order of increasing closure triangle perimeter. Photon noise
contributes \textit{baseline-independent} error.

\subsection{Flat field error} 
\begin{figure} \centering 
\includegraphics[scale=0.3]{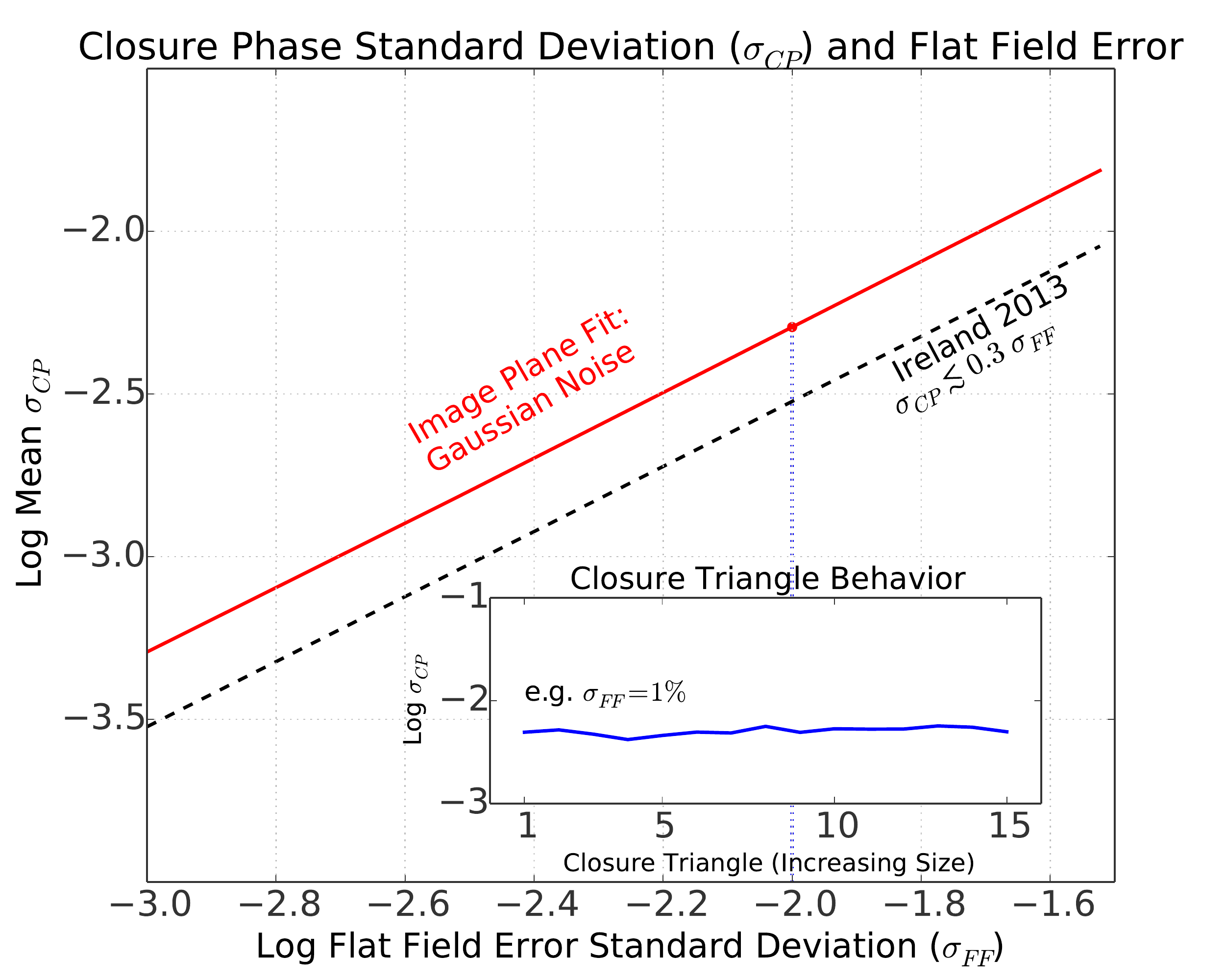}
\caption{\label{fig:gauss}\small  \textbf{Flat field error (monochromatic):}
$\sigma_{CP}$, averaged over closure triangles, is compared to varying
uncertainty in flat field, $\sigma_{FF}$. The solid red line shows our
simulation results.  The dotted black line displays a similar numerical result
from \cite{2013MNRAS.433.1718I}. Uncorrelated flat fielding error induces
\textit{baseline-independent} errors.} 
\end{figure}

We simulated multiplicative flat field error with an arrays of uncorrelated
pixel-to-pixel noise drawn from a Gaussian distribution.  The standard deviation
of the Gaussian distributions range from 0.03\% to 3\%.  \autoref{fig:gauss}
shows our simulation alongside the \cite{2013MNRAS.433.1718I} result:
$$\sigma_{CP} \lesssim 0.3\sigma_F.$$ The small offset between the two results
may be a difference in number of pixels used or in the way error is modeled.
Both results show the same trend.  Flat field error is a
\textit{baseline-independent} effect.

\subsection{Pixel-to-pixel variations in  intra-pixel sensitivity}
\label{sec:ips}

\begin{figure*} \centering \includegraphics[scale=0.6]{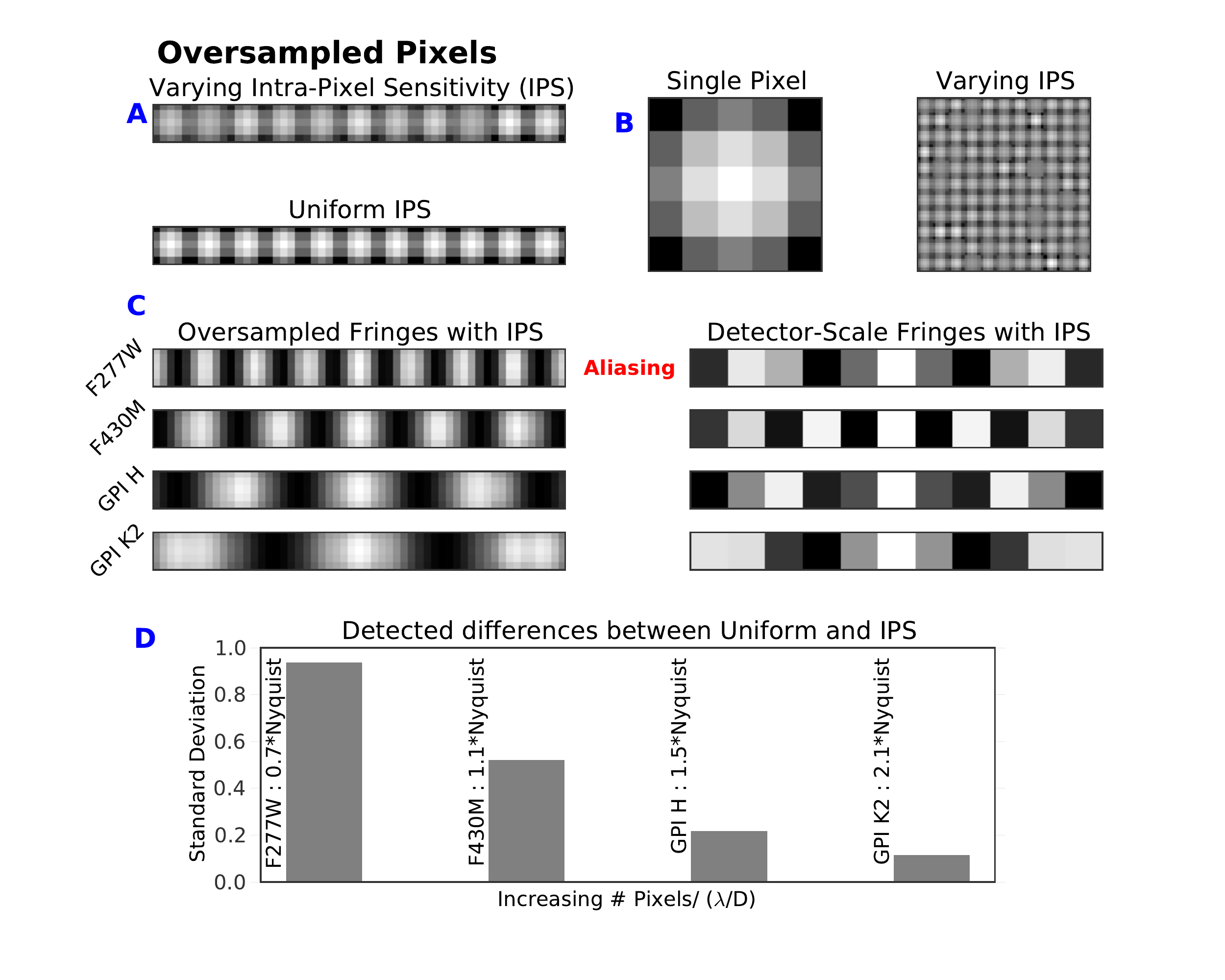}
\caption{\small \textbf{Intra-pixel sensitivity:}  A comparison between uniform
IPS and pixel-to-pixel IPS variations \citep{2008SPIE.7021E..70H} on detectors
with sub- and super-Nyquist pixellation.
\textbf{A.} Top left panel shows the sub-pixel sensitivity variation of two
rows of 11 pixels, one with uniform IPS and the other with IPS realizations
drawn from the statistical model.
\textbf{B.} A single pixel whose sensitivity drops quadratically to 80\% at its
corners, and an oversampled map of $11 \times 11$ pixels drawns from the
statistical IPS model.
\textbf{C.} On the left, a finely sampled fringes ($5\times$ finer than the
detector sampling) with varying IPS. On the right, the same response binned to
the detector scale. At 2.77 $\micron$ the sampling is too coarse to detect the
fringe peaks, which are aliased --- only one peak is visible,
though there are actually three.
\textbf{D.} The difference between image pixel counts for simulated detectors
with uniform pixels and varying IPS.  F430M (just Nyquist sampled) and F277W
(about half Nyquist) show the largest errors. The two well-sampled GPI H and K2
bandpasses show much smaller errors.  \autoref{fig:pixelweight} quantifies
closure phase errors in these situations.} \label{fig:ips} \end{figure*}

We use a data-based model of the NIRISS detector's intra-pixel sensitivity
\citep{2008SPIE.7021E..70H}.  The pixel has maximum sensitivity at its center,
but the sensitivity drops smoothly to $80\% \pm 5\%$ of its peak at the pixel
corners (\autoref{fig:ips}B--left).  We implement a parabolic drop-off of
sensitivity within a pixel.

\cite{1999PASP..111.1434L} describes a single image (integrated over each
pixel) on a detector as
\begin{equation} \label{eqn:pixelresponse} I(x,y)=O(x,y)* P(x,y)
(\mathrm{III}(x,y) * \mathfrak{R}(x,y)) \end{equation}
where the image is a convolution of the object, $O(x,y)$ and the PSF $P(x,y)$
multiplied by a sampling function convolved with the intra-pixel response,
\Rxy.  If \Rxy\  is symmetric it will not contribute phase to the transform of
the image.  However, the intra-pixel sensitivity is not the same for all
pixels, and/or is not symmetric, so is likely to contribute fringe phase error.

Uncharacterized IPS variations are prone to have a larger effect on coarsely
sampled images than finely sampled ones.  \autoref{fig:ips} compares the
effects of sampling frequency when IPS varies from pixel to pixel.  We compare
the sampling in NIRISS F277W and F430M bands, and GPI H and K bands as
illustrative examples. 
 
\begin{figure} \centering
\includegraphics[scale=0.3]{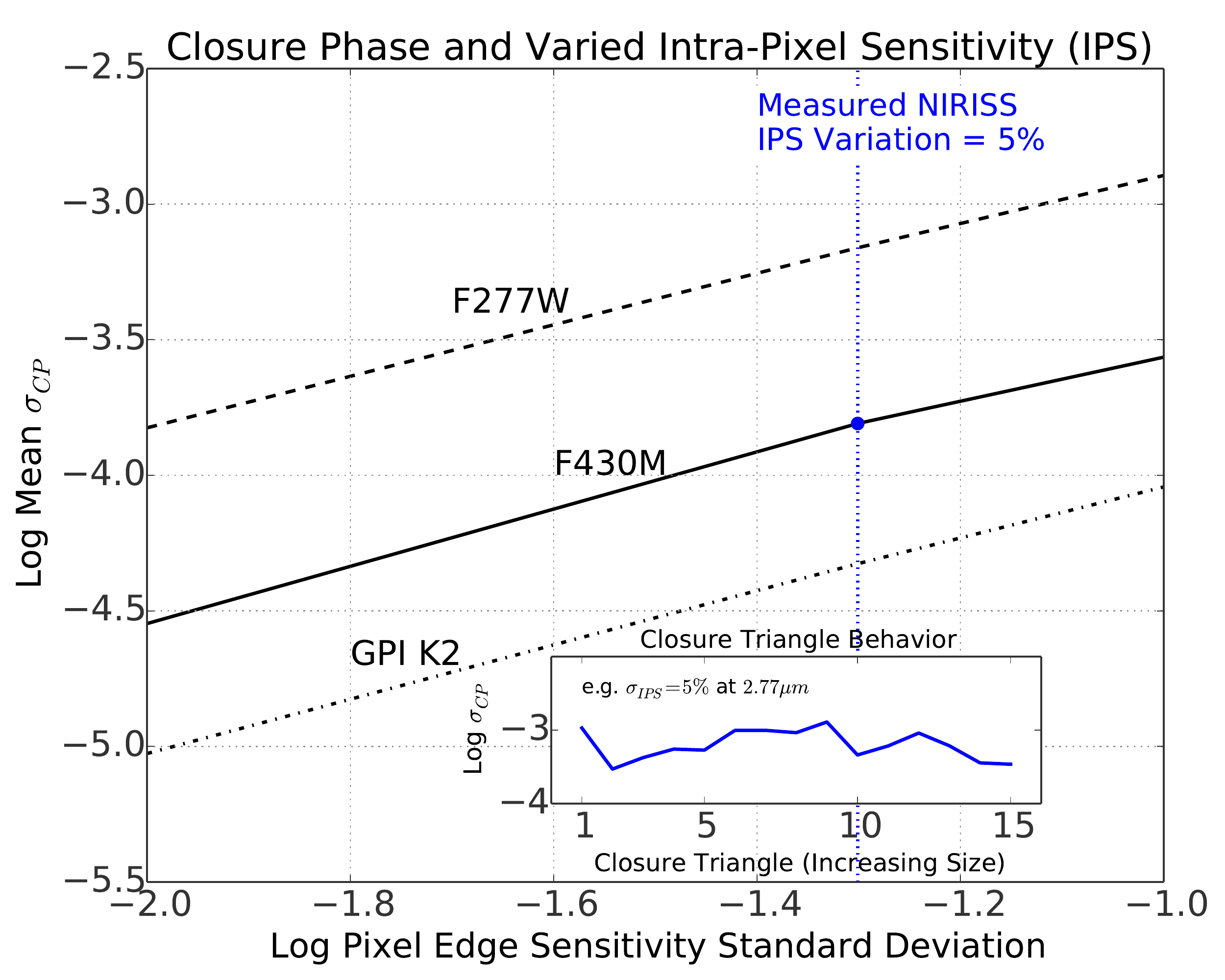}
\caption{\small \textbf{Varied intra-pixel sensitivity (monochromatic):}
Closure phase errors resulting from a range of distributions of IPS.  The
relative edge sensitivity is varied, while maintaining a uniform net pixel
quantum efficiency (see text).  NIRISS detector sampling at 2.77 $\micron$ and
4.3 $\micron$, and GPI sampling at K2 (2.3 $\micron$) are shown.}
\label{fig:pixelweight}
\end{figure}

Here we assume that flat fields are known perfectly, so we use uniform and
symmetric pixel-to-pixel weighting in our model (\autoref{fig:ips}B--left)
and fit data with many realizations of IPS drawn from the model we describe
above  (\eg\ \autoref{fig:ips}B--right).  We renormalize the total pixel
efficiency to maintain a constant net sensitivity of each pixel to avoid
confounding flat-fielding error with IPS effects.  \autoref{fig:pixelweight}
shows increasing closure phase error for increased coarseness in pixel scale.
Although NIRISS' F277W suffers most from IPS variation, we can still achieve
below $10^{-3}$ radians in closure phase error with the 5\% uncertainty in IPS.
Fine scale dithering \citep{JWST.STScI.000647Koe} and careful individual pixel
IPS calibration could mitigate our sensitivity to worse-than-Nyquist pixel
scales.  We note in passing that the NIRISS F277W filter presents an
interesting science opportunity for faint companion imaging.  F277W covers an
$\mathrm{H_2O}$ absorption feature that could help constrain planet chemistry
and dust grain models \citep{2006ApJ...648..614C,2009ApJ...702..154S,
2013arXiv1306.0610C}.

IPS modeling can also help determine fractional-pixel positioning of objects on the
detector, by cross-correlating image data with analytically generated reference
PSFs at different sub-pixel centerings  \citep{2013SPIE.88641L.56}.

% scale=0.6 before double space

\section{PSF magnification and spectral throughput}
\label{sec:polychromatic}
Uncertainty in the coordinate scaling of the PSF affects our linear fit.
Scaling errors could arise from hole size or central wavelength uncertainty.
Coordinate scaling magnifies the PSF envelope and contributes errors, particularly
to longer baselines.  The PSF magnification in the data can be
determined by cross-correlating the power spectrum of the data with power
spectra of model PSFs created using a range of pupil scales.  An NRM can be
used to determine plate scale of IFS data cubes, thus providing an independent
check of either plate scale or wavelength calibration in the hyperspectral data
cubes \citep{2013SPIE.88641V.66}.

Mask geometry or mask scale uncertainties will contribute errors in the
closure phase when there are static pistons. Without static phase errors (\ie\
with a symmetric PSF) there should be no error in closure phase. An asymmetric
PSF (resulting from static piston error), however, will produce
\textit{baseline-dependent} closure phase errors, even when the model perfectly
matches the data.  We demonstrate that random static phases produce
baseline-dependent closure phase error when fitting a polychromatic image,
regardless of whether the model is monochromatic or polychromatic.  Below about
$\lambda/10$ waves (at the central wavelength) rms piston error the closure phase
error scales with piston error. The effect is negligible when piston is
below small fractions of a wave, when the PSF is sufficiently symmetric.  In
this section we use pistons that are uniformly distributed with an rms piston
of 0.06 waves and zero mean (\autoref{tab:simpistons}).

\subsection{Hole size tolerance}
\label{sec:holesize}

\begin{figure}
	\centering
	\includegraphics[scale=0.3]{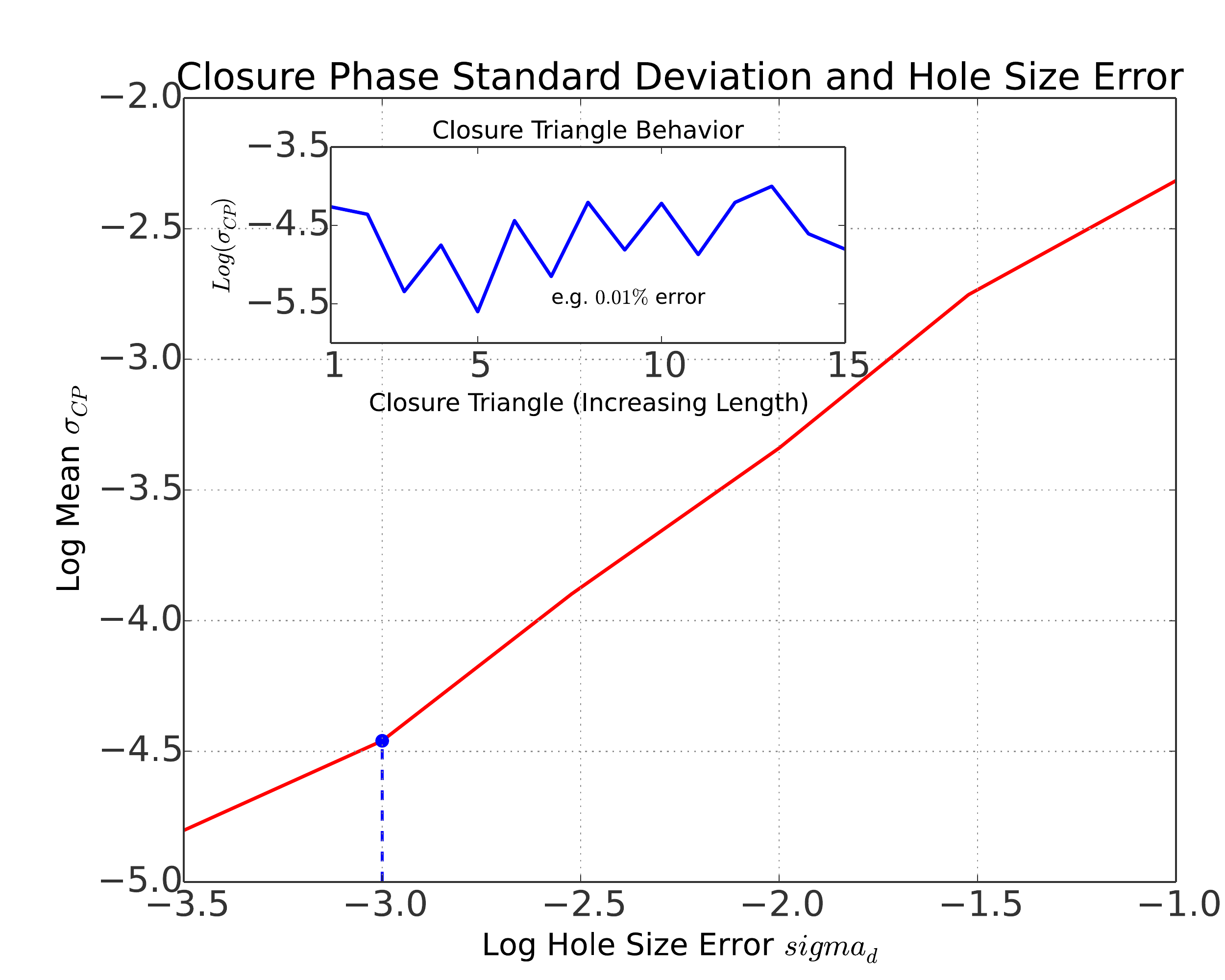}
	\caption{\small \textbf{Hole size errors (monochromatic):}
	We calculate $\sigma_{CP}$ with a range of errors in hole diameter. Hole size
	only affects PSF envelope scaling, which is easily measured in the
	Fourier plane. Wavelength scales the entire PSF.}
	\label{fig:holesize}
	\end{figure}

The mask geometry must be known well to fit the diffractive image model to
data. We simulate masks with 7 identical holes. We used a Gaussian distribution of 
mask hole sizes with varied uncertainties and mean hole diameter $d$, the NRM
hole size for NIRISS. \autoref{fig:holesize} compares hole size error
$\sigma_d$ to $\sigma_{CP}$.  Contrast drops steeply with uncertainty in hole
size in the presence of piston errors.  

\subsection{Fitting medium- and wide-band data}

\begin{figure}
	\centering
	\includegraphics[scale=0.4]{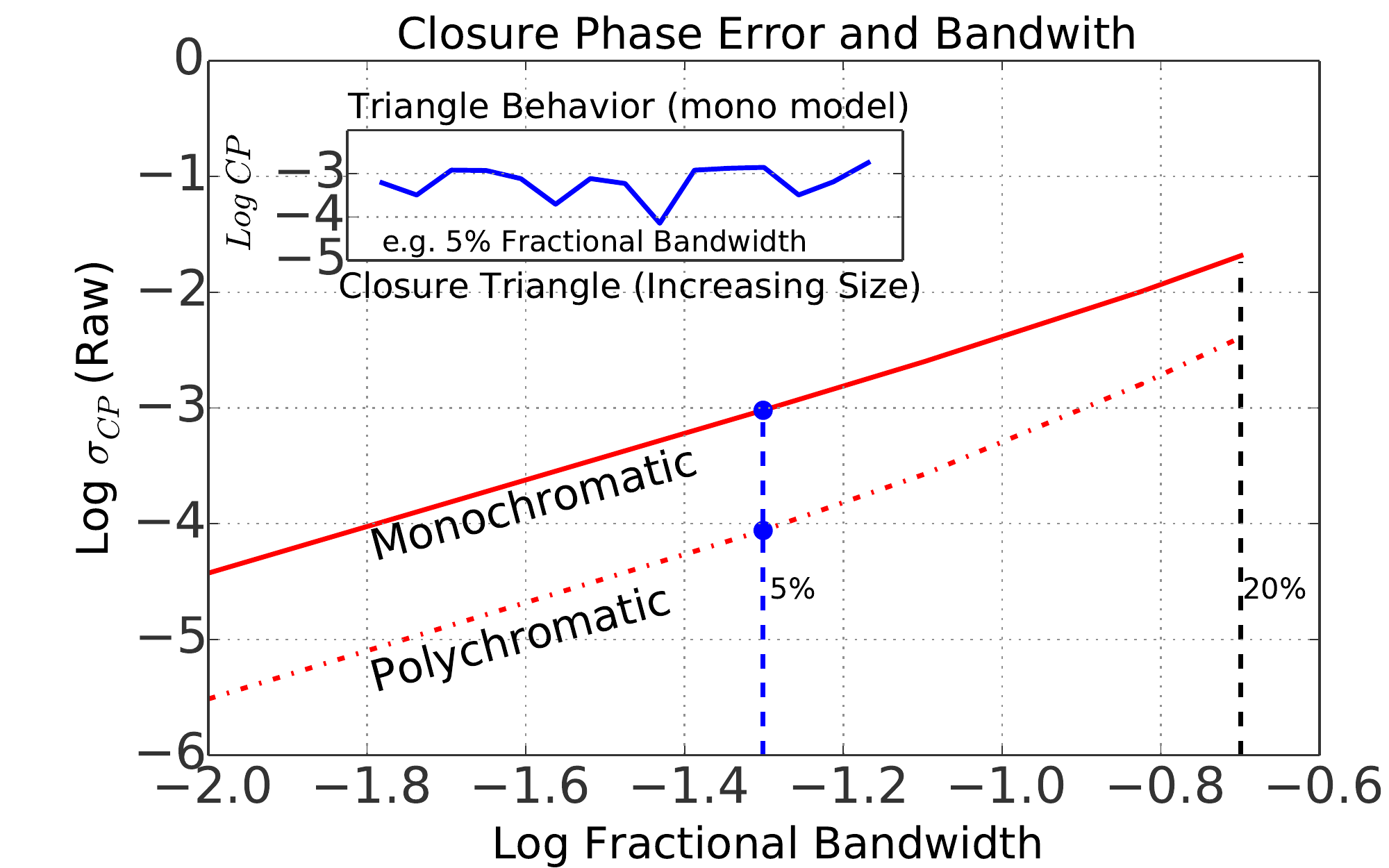}
	\caption{\small \textbf{Fitting finite bandwidth data:} Using a monochromatic
	model to fit finite bandwidth data achieves our required contrast only when the
	fractional bandwidth is $\lesssim$ 1\%.  Using a polychromatic model improves
	contrast by an order of magnitude.  These closure phase errors are
	\textit{baseline-dependent}, and are highly sensitive to the particular
	configuration of holes and pistons.}
	\label{fig:bandpass}
	\end{figure}

NIRISS has three medium-band ($5-8\%$ fractional bandwidth) filters intended
for exoplanet science with NRM. Additionally, the wide-band  F277W filter
($25\%$ fractional bandwidth) may also be scientifically interesting, despite
its coarse sampling.  \autoref{fig:bandpass} demonstrates that a
polychromatic model matches the data better than a monochromatic model,  by
about an order of magnitude.  Closure phase errors for finite-bandwidth images
with pistons are highly \textit{baseline-dependent}. The errors vary by orders
of magnitude between different closure triangles.  Closure phase errors also
depend on the shape of the bandpass for a given realization of piston errors,
when fitting with a polychromatic model.  \autoref{fig:bandpass} suggests
that mismatch in central wavelength can introduce large errors in closure
phase.

\begin{figure}
	\centering
	\includegraphics[scale=0.35]{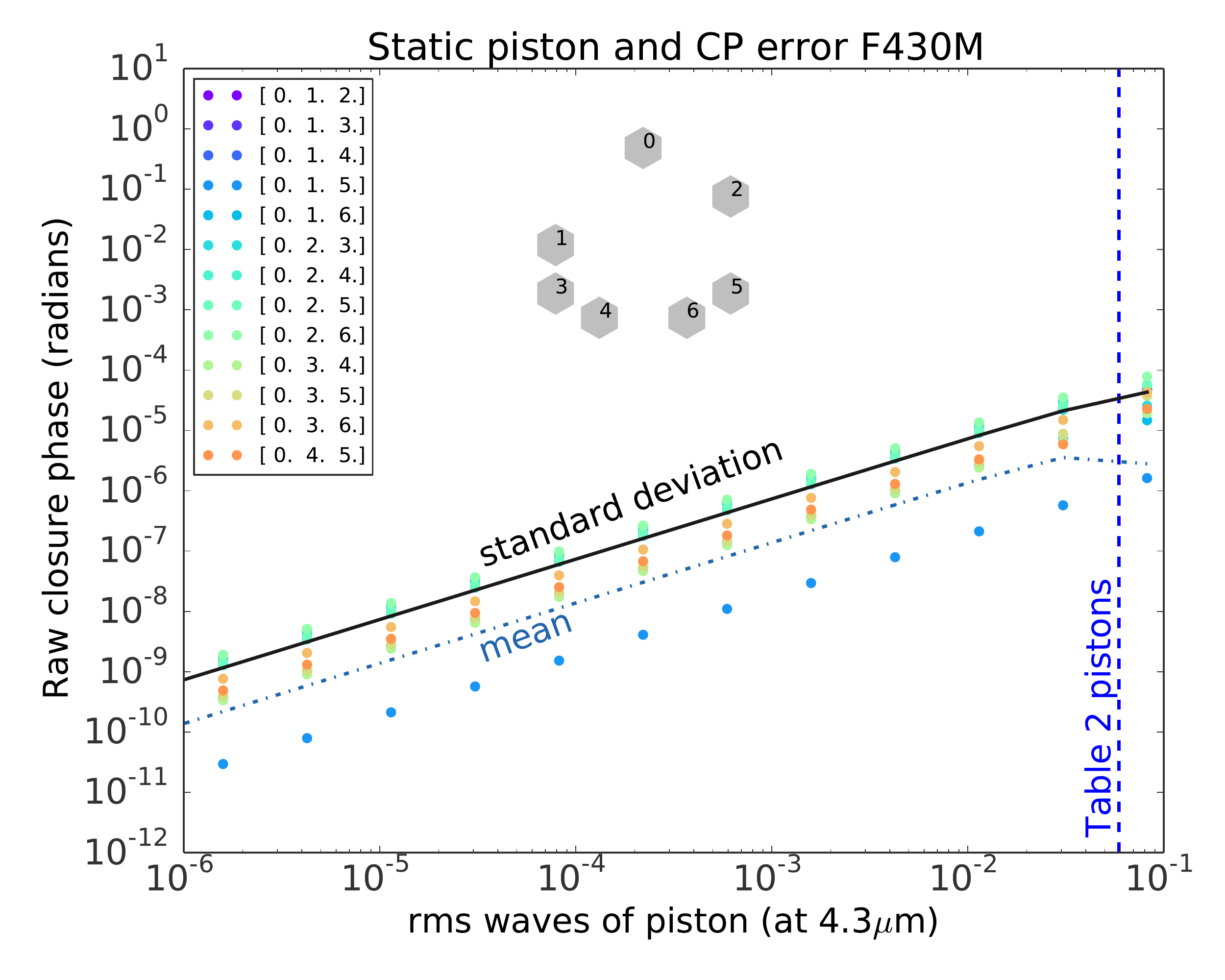}
	\caption{\small \textbf{Spectrum errors and static piston in the pupil:} We
	measure closure phase for polychromatic data with different levels of piston
	simulated with NIRISS's F430M filter and fit with a polychromatic model.  The
	dependence is roughly linear when the pistons are below $\lambda/10$ at the
	central wavelength $\lambda_c = \mathrm{4.3}\micron$.}
	\label{fig:pistonmag}
	\end{figure}

Unless otherwise noted, we use the pistons described in 
\autoref{tab:simpistons} in these simulations.  When there are no pistons in the
pupil, error in the source spectrum should not contribute closure phase errors
because the PSF remains symmetric. To explore how much piston affects closure
phase error for polychromatic data fit with the model described in 
\autoref{eqn:polymodel}, we simulated data with piston at small fractions of a wave
and measured the resulting closure phase. The set of pistons in
\autoref{tab:simpistons} were scaled uniformly to preserve the character of
the errors while changing the size of the error.  Closure phase error scales
with the level of static piston up until an rms piston of about $\lambda/10$ 
(\autoref{fig:pistonmag}).

	\begin{figure}[htbp!]
	\centering
	\includegraphics[scale=0.4]{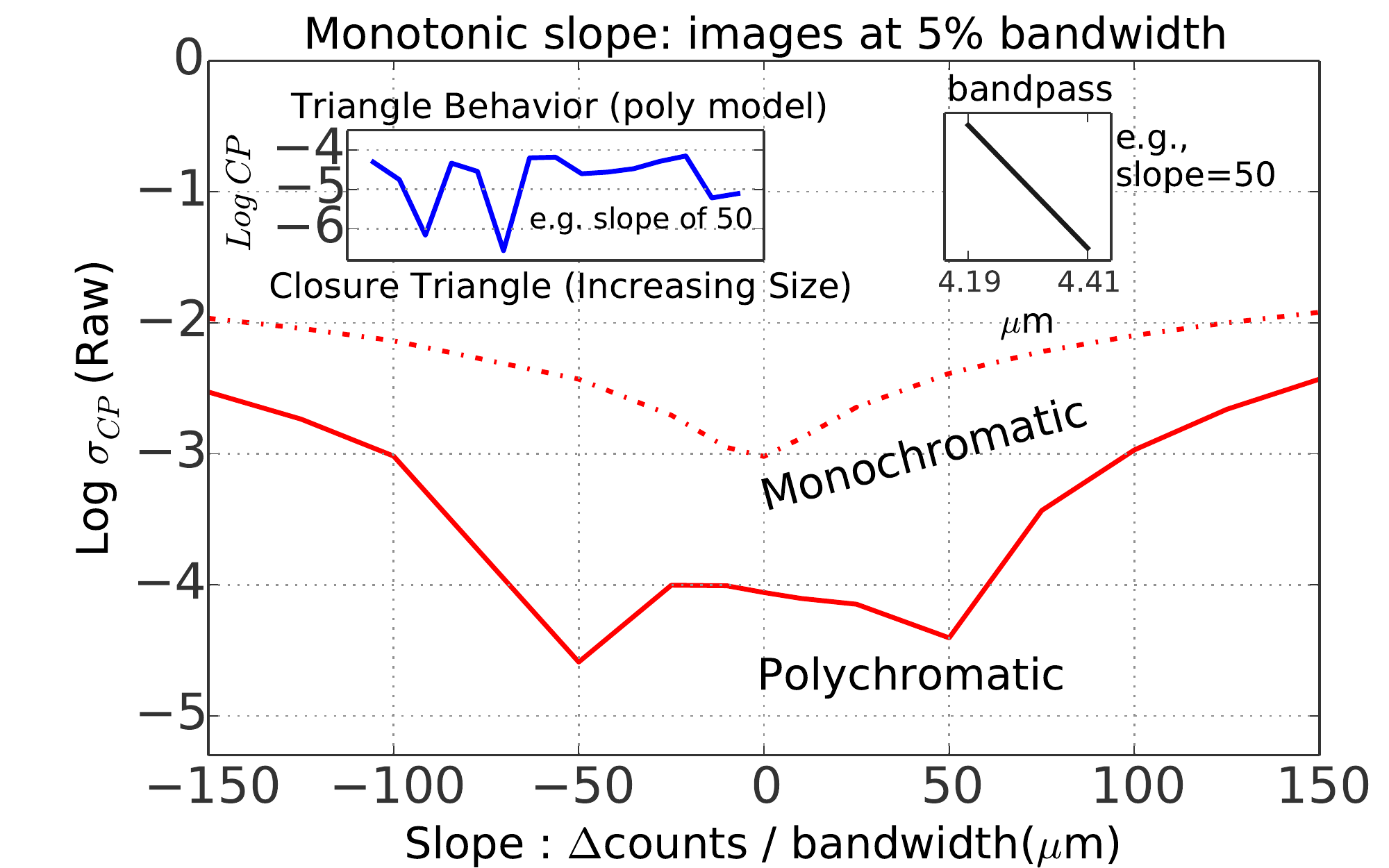}
	\caption{\small \textbf{Bandpass shape:} We fit noiseless polychromatic data
	generated with bandpasses of positive or negative slopes (difference 
	in counts between the edges of the bandpass) with both a
	monochromatic model ($4.3\micron$), and a polychromatic model ($\lambda_c =
	4.3\micron$) matching the bandpass.  Surprisingly, the data are not fit best at
	zero slope with a polychromatic model. The shape of the bandpass can add small
	but significant errors because of averaging discussed in \autoref{sec:fit}.
	Bandpass shape introduces \textit{baseline-dependent} errors. }
	\label{fig:shape}
	\end{figure}

% scale=0.7 before double space
	\begin{figure*}[htbp!]
	\centering
	\includegraphics[scale=0.6]{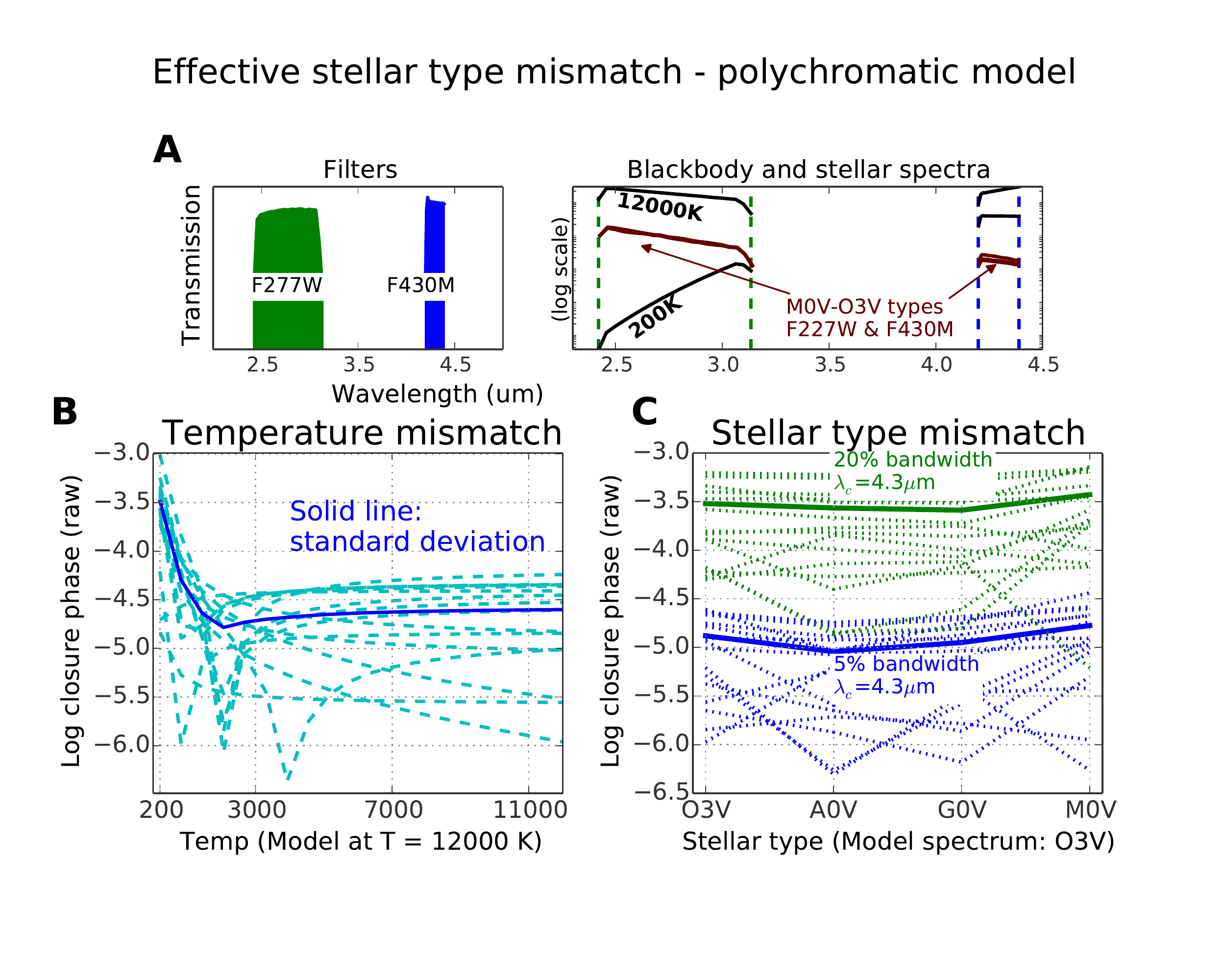}
	\caption{\small \textbf{Spectrum error (polychromatic):}
	\textbf{A.} Left: Transmission profiles for F430M and F277W NIRISS filters. 
	Right: Spectral profiles seen through the two NIRISS filters: blackbody spectra from 
	$T_{eff}$=12000K to $T_{eff}$=200K (subplot \textit{B}) and stellar-type spectra 
	\citep{2006yCat..34540333C} from M0V to O3V (subplot \textit{C}). Large changes
	in spectral type do not strongly affect the slope of the spectrum.
	\textbf{B.} The closure phase error (standard deviation over triangles in
	solid) plotted against blackbody temperature of simulated polychromatic data.
	The model temperature is $T_{eff}$=12000K. Dashed lines represent individual 
	closure triangles.
	\textbf{C.} Closure phase is plotted against mismatch in stellar-type between the
	model created with an O3V stellar spectrum at a central wavelength
	$\lambda_c$=4.3$\mu$m for 5\% (blue) and 20\% (green) bandpasses. Closure
	phases for all of 15 triangles are in dotted lines and their standard deviation
	in solid.} \label{fig:polymodel}
	\end{figure*}

Fitting polychromatic data with a polychromatic model will only introduce
closure phase errors when there are non-zero pistons in the pupil and the PSF
has asymmetries. The gross characteristics of the net throughput (filter
$\times$ source spectrum) sets a floor on raw contrast (\autoref{fig:shape}).
This is likely because of the piston averaging mentioned in section
\autoref{sec:fit}. Polychromatic fitting is robust to smaller differences in
throughput (e.g. slope) between the model and data.

JWST is anticipated to have 80\% Strehl Ratio at 2 \micron. If its all wavefront
error is in piston, piston standard deviation should be about $0.035$ waves
at 4.3 \micron. These values fall below $\lambda/10$ for the F430M filter.  We
assume that the filter is well known but the source spectrum is not.  We
investigate mismatch in blackbody spectrum (\ie\ temperature) as well as an
incorrect choice of stellar spectral type. We generate point source images for
NIRISS F430M and F277W at a range of temperatures and stellar spectra, and fit
each to either a $T=12000K$ blackbody or an O3V star model.

\autoref{fig:polymodel}A displays the F430M and F277W filter transmissions
and range of blackbody curves, from our modeled spectrum at T=12000K down to
T=200K.  \autoref{fig:polymodel}B displays raw closure phases for all of the 15
independent baselines and their standard deviation from fitting an incorrect
blackbody model (T=12000K) to data simulated at our range of temperatures.
Similar to the bandpass shape simulation in \autoref{fig:shape}, there is an
overall throughput shape that yields best sensitivity, though this behavior
varies depending on baseline. The polychromatic image plane model is robust to
a large error in blackbody temperature, up until the blackbody slope turns over
for extremely cool objects (\eg cooler than ~670K at 4.3 \micron).

We repeat the same procedure with a range of stellar photosphere models from
\cite{2006yCat..34540333C} to examine the effect of stellar type mismatch on
closure phase sensitivity.  \autoref{fig:polymodel}A shows that the range of
stellar spectra from O3V to M0V stars in the two JWST filters do no differ much
by slope. We model a flat bandpass centered at $\lambda_c$=4.3$\mu$m at 5\% and
20\% to introduce the same rms piston.  \autoref{fig:polymodel}C shows the
raw closure phase error when the two models (using stellar type O3V) are fit to
this range of spectra. The polychromatic image plane model shows similar
robustness to poor knowledge of the stellar spectral type.

Polychromatic model fitting will not be a limiting factor on raw contrast for
the narrowest filters with less than 0.06 waves rms piston, but becomes more of
a concern for wide-band images, especially at shorter wavelengths.  For small
piston WFE, raw contrast for NRM with the F277W filter is not limited by the
size of the bandpass compared to the effect of intrapixel sensitivity or flat
field errors.

\section{Higher Spatial Frequency Wavefront Error}
\label{sec:WFE}

Wavefront errors can introduce both amplitude and phase aberration in an image.
NRM is most effective in the low spatial frequency wavefront error regime;
closure phases are insensitive, to first and second order, to piston wavefront
error \citep{2013MNRAS.433.1718I}. Wavefront errors on mirror segments often
span a range of spatial frequencies. 
We used  WebbPSF \citep{2012SPIE.8442E..3DP} to simulate NIRISS NRM PSFs with
low to mid spatial frequency wavefront errors, including segment tip and tilt,
and figure errors on the segments and instrument optics.

\begin{figure}[htbp!] \centering \includegraphics[scale=0.3]{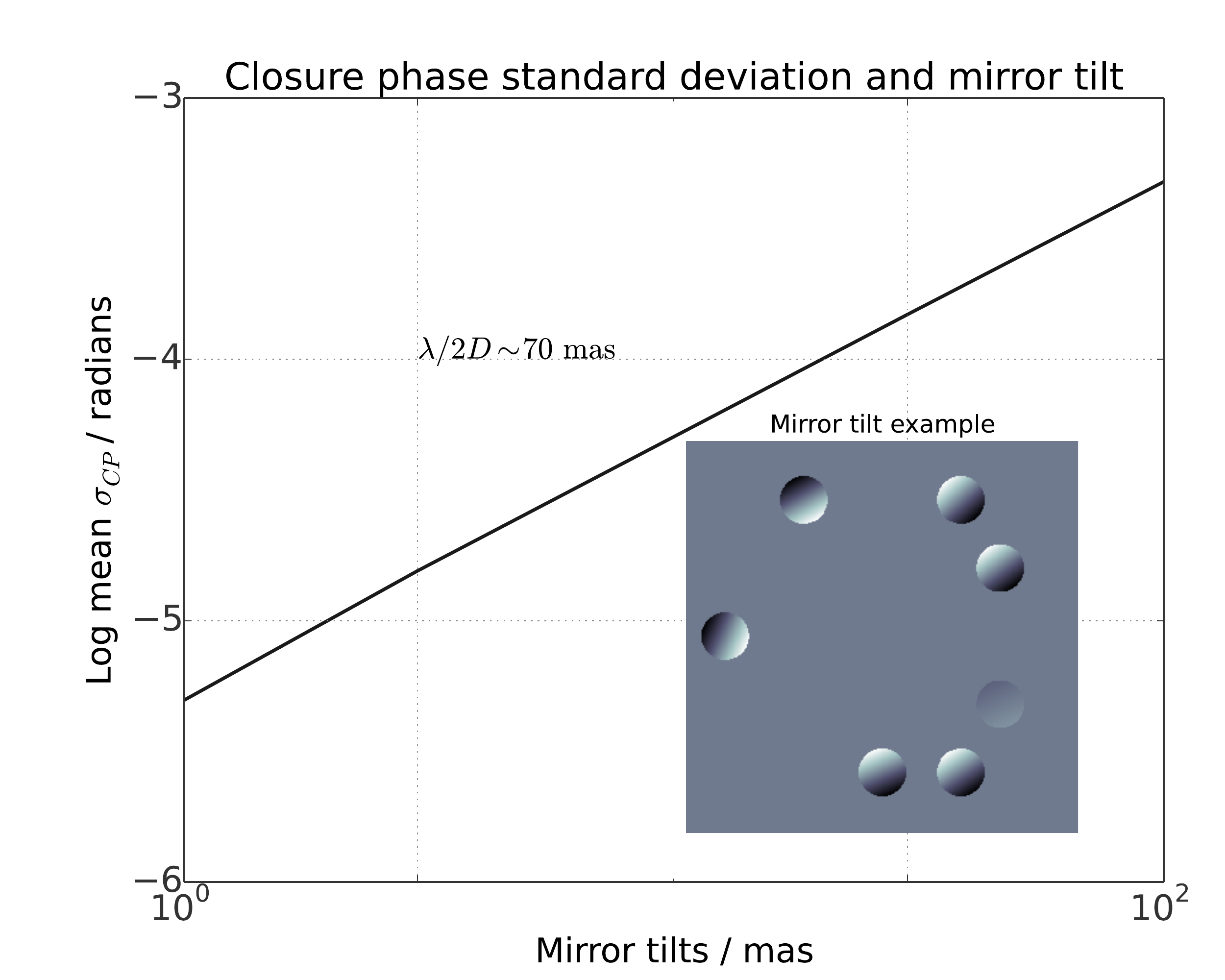}
	\includegraphics[scale=0.3]{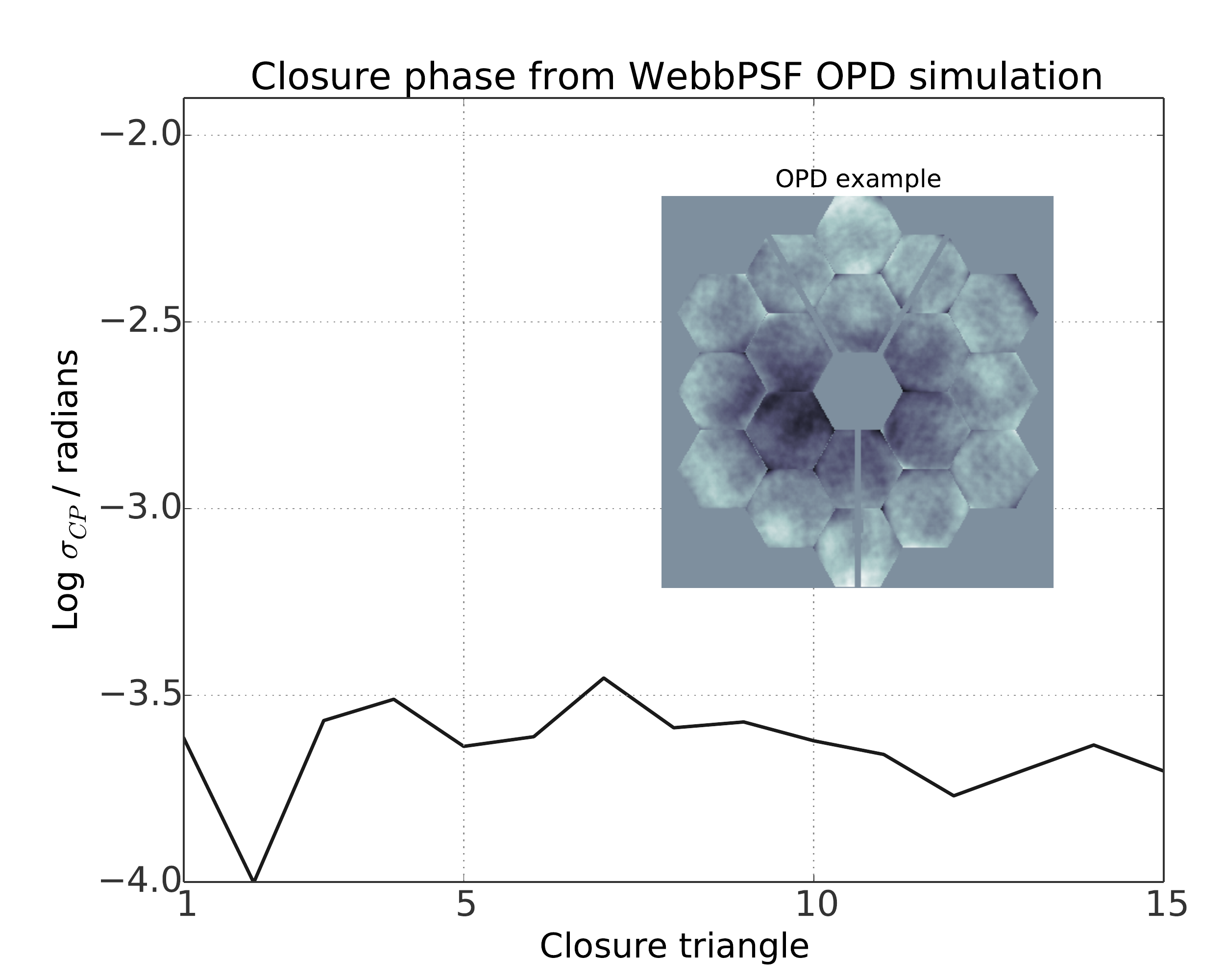} \caption{\small \label{fig:WFE}
	\textbf{Tilts and higher order wavefront error:} \textit{Top:} We measure
	$\sigma_{CP}$ from 100 different tilt error simulations of various sizes of
	tilt.  An instance of tilt over each hole is inset in the top panel.
	\textit{Bottom:} $\sigma_{CP}$ from fitting 10 different $\sim140~\mathrm{nm}$
	rms JWST NIRISS wavefront realizations containing higher order wavefront error
	\citep{2012SPIE.8449E..0VK}. PSFs were generated with WebbPSF software
	\citep{2012SPIE.8442E..3DP}.} \end{figure}

We first explore the contribution of tip/tilt error by introducing randomly
oriented tilts on simulated JWST mirrors. In \autoref{fig:WFE} each trial
has a fixed tilt magnitude, which we place on 6 mirrors.  We require a mean tilt of
zero by constraining the last mirror so that we do not actually shift the
centering of the PSF.  Closure phase errors of $10^{-4}$ result from segment
tilts of the order of half a resolution element.

We also calculated closure phases for 10 different PSF realizations of $\sim140
~\mathrm{nm}$ rms optical path delay (OPD) on the JWST primary
\citep{2012SPIE.8449E..0VK} using WebbPSF (\autoref{fig:WFE}).  These sample
OPDs do not contain significant segment tilt or global focus. Fitting simulated
data with this WFE yields $\sigma_{CP} =10^{-3.5}$.

\section{Detecting the companion around \texorpdfstring{L\MakeLowercase{k}C\MakeLowercase{a}15 }\ with JWST NIRISS}
%\section{Detecting the companion around LkCa15 with JWST NIRISS}
%
	\begin{figure}[htbp!] 
	\centering
	\label{fig:lkca} 
	\includegraphics[scale=0.35]{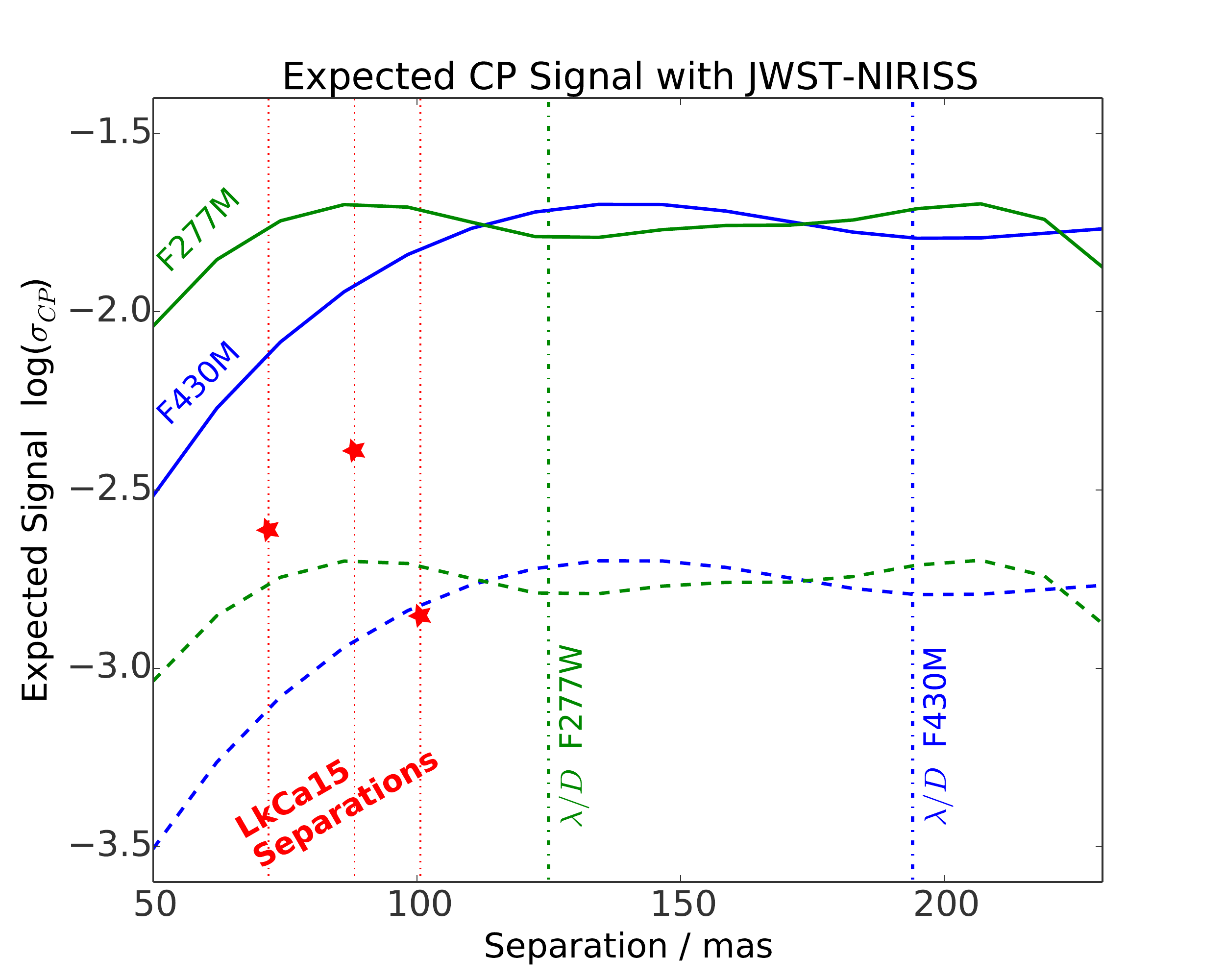}
	\caption{\small \label{fig:LkCa15} \textbf{LkCa15 with JWST NIRISS:} The
	theoretical closure phases for a single binary companion predict a required
	contrast $\approx1/\sigma_{CP} $ empirically. The plot shows $\sigma_{CP}$ over
	all possible triangles through NIRISS F277W and F430M filters. If F277W
	observations can reliably measure closure phase below $10^{-3}$, they could
	detect the $K'$ signal of the companion structure around LkCa15, according to
	\cite{2012ApJ...745....5K}. Two of their three companion sources lie between
	$\lambda/2D$ and $\lambda/D$ at 2.77 \micron. }
	\end{figure}

We consider the case of LkCa 15, which has a detected companion inside its disk
gap, to compare simulated NIRISS NRM to current ground-based NRM. In 
\autoref{fig:LkCa15} we plot the theoretical binary closure phase signal for the
mask on NIRISS at $4.3 \ \micron$ and $2.77 \micron$  for two flux ratios:
$10^{-2}$ and $10^{-3}$.  The red stars mark the Keck Telescope $K'$ and $L'$
detections of companion structure around LkCa15, a transitional disk with
potential planet-forming bodies \citep{2012ApJ...745....5K}. The LkCa15
detections fall between $0.5 \lambda/D$ and $\lambda/D$ for $2.77 \ \micron$ (D
referring to longest baseline), and are detectable with contrast better than
$10^{-3}$.

The F277W filter, at shorter wavelength, will access smaller inner working
angles, relevant to following up close companions detected with large apertures
on the ground. Following our analysis in
\autoref{sec:data}--\autoref{sec:WFE}, we conclude that F277W would be able to achieve
raw contrast of $10^{-3}$ with good flat field measurements and IPS
characterization. With this performance NIRISS would be able to detect the
LkCa15 companion signal in a routine observation.

At 2.77\micron, \ $\lambda/2B\sim\mathrm{88}\ \mathrm{mas}$ for NIRISS NRM.  Detections
this close to the diffraction limit maybe challenging from a modeling
perspective.  The anticipated stability of NIRISS NRM's visibility amplitudes
may help in breaking contrast-separation degeneracies more securely than in
similar ground-based data.  Multi-band observations may also help resolve
degeneracies between separation and contrast. While the angular separation
LkCa15's companions may still pose a challenge, NIRISS will likely achieve
better contrast than ground-based NRM.

\section{Discussion and Recommendations} 

Our analysis of the limits on raw contrast helps develop instrument
tests, calibration needs, and observing strategies for JWST NIRISS.  We applied
our analytic model to NIRISS' cryogenic test data \citep{AlexSPIE2014}.  The
necessary conditions for $10^{-4}$ closure phase error with NIRISS NRM depend
strongly on detector behavior and other instrument and telescope details.
Instrument characterization and accurate pointing and dithering, coupled with
point source calibration of science target data could help achieve this
contrast goal for routine NIRISS observations.  Because of thermal drift and
the planned occasional wavefront control activity in flight
\citep{2004SPIE.5487..825L,2006SSRv..123..485G,2008SPIE.7010E..22M},
near-contemporaneous acquisition of target and point source calibrator data is
desirable.  Since NIRISS point source calibration involves system-wide
complexities we defer study of it here.
NRM paired with the sub-Nyquist sampled F277W filter should provide about 7.5
magnitude raw contrast, which could be useful for probing water absorption
features.

Identifying outlying image pixels is straightforward with our analytical model
fit.  Our algorithm also makes for more efficient observing strategies since
missing pixel data do not need to be filled in with dithered observations.
Pixel-to-pixel variation in sub-pixel scale differences is easily incorporated
in a statistical or detailed manner in our model fit approach, and saturated
pixels can be ignored in the fit.  This is relevant to NIRISS,
with its barely-Nyquist pixel scale and JWST's limited lifetime.

Flat field errors
of 0.1\% limit raw monochromatic contrast to a few $\times 10^{-4}$. Precise
positioning of the target and calibrator on the same pixel will reduce the
effect of flat field errors \citep{2009SPIE.7440E..30S}.  Knowledge of
intra-pixel sensitivity can be used to improve astrometry and reduce fringe
phase measurement errors \citep{2013SPIE.88641L.56}.
Intra-pixel sensitivity (IPS) variations over the detector limit contrast,
especially for the shorter wavelength NIRISS F277W and F380M filters.  However,
knowledge of the IPS can ameliorate this.  We demonstrated that a statistical
understanding of IPS variations can help markedly.  As with flat field errors,
repeated placement of targets to sub-pixel accuracy will benefit NIRISS NRM's
contrast.

Small scaling errors may introduce closure phase errors when there are
static piston in the pupil.  Matching magnification between data and
model is straightforward in the Fourier
domain.  For instruments that have IFSs, this technique can be used either for
wavelength or for plate scale calibration of individual hyperspectral cube
slices \citep{2013SPIE.88641V.66}. 

NIRISS NRM data analysis in the image plane will benefit from polychromatic
modeling.  The necessary piston averaging at the band's central wavelength will
contribute closure phase errors when there are non-zero pistons in the pupil.
However, the modeled bandpass is fairly robust to errors in source spectrum, as
long as the spectrum slope sign is correct.  Raw contrast from F277W
observations will be reduced because of its wider bandpass, and because F277W
will see higher instrument WFE at its shorter wavelength. However, with the
anticipated WFE for JWST, its bandpass should not be the limiting factor for
contrast. 

The primary contrast limiting factors are pixel-to-pixel (flat field)
variations and IPS variation for the coarsest sampled F277W filter. In
comparison contrast will be largely unaffectected by uncertainty in source
spectrum if the modeled bandpass roughly matches the data. While these various
systematics limit raw contrast, additional sensitivity will be possible through
point-source calibrations and leveraging stable closure amplitudes.

Flat field errors can also effect closure phase measurements from IFS images.
but higher order wavefront error from atmospheric effects may be the biggest
limiting factor on the ground. These higher order errors certainly exist on
ground-based instruments like GPI, and may contribute amplitude as well
as phase errors.  JWST NIRISS' wavefront error is expected to be dominated by
low order terms, and stay below about 160~nm rms.  Fitting uncorrelated pistons
with our analytic model is robust to low-order wavefront errors including tip
and tilt.  A thorough study of the effects of higher-frequency wavefront error
in the NRM PSF is warranted.

Space-based NRM presents opportunities for extended object imaging at high
angular resolution. Centro-symmetric structures require amplitude measurements,
which will be stable in the absence of atmospheric effects.
Space-based NRM's fringe phase and amplitude measurements provide true imaging,
which can benefit AGN and quasar science \citet{2014ApJ...783...73F}, so our
image plane model could improve observing efficiency and data reduction methods
for space-based high resolution imaging.  An analytic point source model is a
step towards more sophisticated forward-modeling of NRM data.

\acknowledgements We acknowledge Ron Allen, Anthony Cheetham, Erin Elliot,
\'Etienne Artigau, and R\'emi Soummer for useful comments and both anonymous
referees for insightful suggestions. This material is based upon work supported 
in part by the National Science Foundation Graduate Research Fellowship Program under 
Grant No. DGE-1232825, by NASA grant APRA08-0117,
and the STScI Director's Discretionary Research Fund.
\bibliographystyle{apj}
%\bibliography{fringes}

\begin{thebibliography}{}
\expandafter\ifx\csname natexlab\endcsname\relax\def\natexlab#1{#1}\fi

\bibitem[{{Baldwin} {et~al.}(1986){Baldwin}, {Haniff}, {Mackay}, \&
  {Warner}}]{1986Natur.320..595B}
{Baldwin}, J.~E., {Haniff}, C.~A., {Mackay}, C.~D., \& {Warner}, P.~J. 1986,
  \nat, 320, 595

\bibitem[{{Beichman} {et~al.}(2010){Beichman}, {Krist}, {Trauger}, {Greene},
  {Oppenheimer}, {Sivaramakrishnan}, {Doyon}, {Boccaletti}, {Barman}, \&
  {Rieke}}]{2010PASP..122..162B}
{Beichman}, C.~A., {Krist}, J., {Trauger}, J.~T., {et~al.} 2010, \pasp, 122,
  162

\bibitem[{{Bernat} {et~al.}(2010){Bernat}, {Bouchez}, {Ireland}, {Tuthill},
  {Martinache}, {Angione}, {Burruss}, {Cromer}, {Dekany}, {Guiwits}, {Henning},
  {Hickey}, {Kibblewhite}, {McKenna}, {Moore}, {Petrie}, {Roberts}, {Shelton},
  {Thicksten}, {Trinh}, {Tripathi}, {Troy}, {Truong}, {Velur}, \&
  {Lloyd}}]{2010ApJ...715..724B}
{Bernat}, D., {Bouchez}, A.~H., {Ireland}, M., {et~al.} 2010, \apj, 715, 724

\bibitem[{{Beuzit} {et~al.}(2010){Beuzit}, {Feldt}, {Mouillet}, {Dohlen},
  {Puget}, {Wildi}, \& {SPHERE Consortium}}]{2010lyot.confE..44B}
{Beuzit}, J.-L., {Feldt}, M., {Mouillet}, D., {et~al.} 2010, in In the Spirit
  of Lyot 2010

\bibitem[{{Beuzit} {et~al.}(2008){Beuzit}, {Feldt}, {Dohlen}, {Mouillet},
  {Puget}, {Wildi}, {Abe}, {Antichi}, {Baruffolo}, {Baudoz}, {Boccaletti},
  {Carbillet}, {Charton}, {Claudi}, {Downing}, {Fabron}, {Feautrier},
  {Fedrigo}, {Fusco}, {Gach}, {Gratton}, {Henning}, {Hubin}, {Joos}, {Kasper},
  {Langlois}, {Lenzen}, {Moutou}, {Pavlov}, {Petit}, {Pragt}, {Rabou}, {Rigal},
  {Roelfsema}, {Rousset}, {Saisse}, {Schmid}, {Stadler}, {Thalmann}, {Turatto},
  {Udry}, {Vakili}, \& {Waters}}]{2008SPIE.7014E..41B}
{Beuzit}, J.-L., {Feldt}, M., {Dohlen}, K., {et~al.} 2008, in Society of
  Photo-Optical Instrumentation Engineers (SPIE) Conference Series, Vol. 7014

\bibitem[{{Biller} {et~al.}(2013){Biller}, {Liu}, {Wahhaj}, {Nielsen},
  {Hayward}, {Males}, {Skemer}, {Close}, {Chun}, {Ftaclas}, {Clarke}, {Thatte},
  {Shkolnik}, {Reid}, {Hartung}, {Boss}, {Lin}, {Alencar}, {de Gouveia Dal
  Pino}, {Gregorio-Hetem}, \& {Toomey}}]{2013ApJ...777..160B}
{Biller}, B.~A., {Liu}, M.~C., {Wahhaj}, Z., {et~al.} 2013, \apj, 777, 160

\bibitem[{{Bos} {et~al.}(2008){Bos}, {Kubalak}, {Antonille}, {Ohl}, {Hagopian},
  {Davila}, {Sullivan}, {S{\'a}nchez}, {Sabatke}, {Woodruff}, {te Plate},
  {Evans}, {Isbrucker}, {Somerstein}, {Wells}, \&
  {Ronayette}}]{2008SPIE.7010E..98B}
{Bos}, B.~J., {Kubalak}, D.~A., {Antonille}, S.~R., {et~al.} 2008, in Society
  of Photo-Optical Instrumentation Engineers (SPIE) Conference Series, Vol.
  7010

\bibitem[{{Castelli} \& {Kurucz}(2006)}]{2006yCat..34540333C}
{Castelli}, F., \& {Kurucz}, R.~L. 2006, VizieR Online Data Catalog, 345, 40333

\bibitem[{Cheetham {et~al.}(2012)Cheetham, Tuthill, Sivaramakrishnan, \&
  Lloyd}]{Cheetham:12}
Cheetham, A.~C., Tuthill, P.~G., Sivaramakrishnan, A., \& Lloyd, J.~P. 2012,
  Opt. Express, 20, 29457

\bibitem[{{Cieza} {et~al.}(2013){Cieza}, {Lacour}, {Schreiber}, {Casassus},
  {Jord{\'a}n}, {Mathews}, {C{\'a}novas}, {M{\'e}nard}, {Kraus}, {P{\'e}rez},
  {Tuthill}, \& {Ireland}}]{2013ApJ...762L..12C}
{Cieza}, L.~A., {Lacour}, S., {Schreiber}, M.~R., {et~al.} 2013, \apjl, 762,
  L12

\bibitem[{{Currie} {et~al.}(2013){Currie}, {Burrows}, {Madhusudhan},
  {Fukagawa}, {Girard}, {Dawson}, {Murray-Clay}, {Kenyon}, {Kuchner},
  {Matsumura}, {Jayawardhana}, {Chambers}, \& {Bromley}}]{2013arXiv1306.0610C}
{Currie}, T., {Burrows}, A., {Madhusudhan}, N., {et~al.} 2013, ArXiv e-prints,
  arXiv:1306.0610

\bibitem[{{Cushing} {et~al.}(2006){Cushing}, {Roellig}, {Marley}, {Saumon},
  {Leggett}, {Kirkpatrick}, {Wilson}, {Sloan}, {Mainzer}, {Van Cleve}, \&
  {Houck}}]{2006ApJ...648..614C}
{Cushing}, M.~C., {Roellig}, T.~L., {Marley}, M.~S., {et~al.} 2006, \apj, 648,
  614

\bibitem[{{Doyon} {et~al.}(2012){Doyon}, {Hutchings}, {Beaulieu}, {Albert},
  {Lafreni{\`e}re}, {Willott}, {Touahri}, {Rowlands}, {Maszkiewicz},
  {Fullerton}, {Volk}, {Martel}, {Chayer}, {Sivaramakrishnan}, {Abraham},
  {Ferrarese}, {Jayawardhana}, {Johnstone}, {Meyer}, {Pipher}, \&
  {Sawicki}}]{2012SPIE.8442E..2RD}
{Doyon}, R., {Hutchings}, J.~B., {Beaulieu}, M., {et~al.} 2012, in Society of
  Photo-Optical Instrumentation Engineers (SPIE) Conference Series, Vol. 8442

\bibitem[{{Ford} {et~al.}(2014){Ford}, {McKernan}, {Sivaramakrishnan},
  {Martel}, {Koekemoer}, {Lafreni{\`e}re}, \&
  {Parmentier}}]{2014ApJ...783...73F}
{Ford}, K.~E.~S., {McKernan}, B., {Sivaramakrishnan}, A., {et~al.} 2014, \apj,
  783, 73

\bibitem[{{Gardner} {et~al.}(2006){Gardner}, {Mather}, {Clampin}, {Doyon},
  {Greenhouse}, {Hammel}, {Hutchings}, {Jakobsen}, {Lilly}, {Long}, {Lunine},
  {McCaughrean}, {Mountain}, {Nella}, {Rieke}, {Rieke}, {Rix}, {Smith},
  {Sonneborn}, {Stiavelli}, {Stockman}, {Windhorst}, \&
  {Wright}}]{2006SSRv..123..485G}
{Gardner}, J.~P., {Mather}, J.~C., {Clampin}, M., {et~al.} 2006, \ssr, 123, 485

\bibitem[{{Girard}(2013)}]{NACO}
{Girard}, J. 2013, NACO User Manual, European Southern Observatory,
  {VLT-MAN-ESO-14200-2761}

\bibitem[{{Greenbaum} {et~al.}(2014){Greenbaum}, {Martel}, {Sivaramakrishnan},
  {Volk}, {Pueyo}, \& {Tuthill}}]{AlexSPIE2014}
{Greenbaum}, A.~Z., {Martel}, A., {Sivaramakrishnan}, A., {et~al.} 2014, in
  Society of Photo-Optical Instrumentation Engineers (SPIE) Conference Series,
  in prep

\bibitem[{{Greenbaum} {et~al.}(2013{\natexlab{a}}){Greenbaum},
  {Sivaramakrishnan}, \& {Pueyo}}]{2013SPIE.88641L.56}
{Greenbaum}, A.~Z., {Sivaramakrishnan}, S., \& {Pueyo}, L. 2013{\natexlab{a}},
  in Society of Photo-Optical Instrumentation Engineers (SPIE) Conference
  Series, Vol. 8864

\bibitem[{{Greenbaum} {et~al.}(2013{\natexlab{b}}){Greenbaum},
  {Sivaramakrishnan}, {Pueyo}, {Ingraham}, {Thomas}, {Wolff}, {Perrin},
  {Norris}, \& {Tuthill}}]{2013SPIE.88641V.66}
{Greenbaum}, A.~Z., {Sivaramakrishnan}, S., {Pueyo}, L., {et~al.}
  2013{\natexlab{b}}, in Society of Photo-Optical Instrumentation Engineers
  (SPIE) Conference Series, Vol. 8864

\bibitem[{{Haniff} {et~al.}(1987){Haniff}, {Mackay}, {Titterington}, {Sivia},
  \& {Baldwin}}]{1987Natur.328..694H}
{Haniff}, C.~A., {Mackay}, C.~D., {Titterington}, D.~J., {Sivia}, D., \&
  {Baldwin}, J.~E. 1987, \nat, 328, 694

\bibitem[{{Hardy} {et~al.}(2008){Hardy}, {Baril}, {Pazder}, \&
  {Stilburn}}]{2008SPIE.7021E..70H}
{Hardy}, T., {Baril}, M.~R., {Pazder}, J., \& {Stilburn}, J.~S. 2008, in
  Society of Photo-Optical Instrumentation Engineers (SPIE) Conference Series,
  Vol. 7021

\bibitem[{{Hinkley} {et~al.}(2011){Hinkley}, {Carpenter}, {Ireland}, \&
  {Kraus}}]{2011ApJ...730L..21H}
{Hinkley}, S., {Carpenter}, J.~M., {Ireland}, M.~J., \& {Kraus}, A.~L. 2011,
  \apjl, 730, L21

\bibitem[{{Hu{\'e}lamo} {et~al.}(2011){Hu{\'e}lamo}, {Lacour}, {Tuthill},
  {Ireland}, {Kraus}, \& {Chauvin}}]{2011AA...528L...7H}
{Hu{\'e}lamo}, N., {Lacour}, S., {Tuthill}, P., {et~al.} 2011, \aap, 528, L7

\bibitem[{{Ireland}(2013)}]{2013MNRAS.433.1718I}
{Ireland}, M.~J. 2013, \mnras, 433, 1718

\bibitem[{{Ireland} {et~al.}(2011){Ireland}, {Kraus}, {Martinache}, {Law}, \&
  {Hillenbrand}}]{2011ApJ...726..113I}
{Ireland}, M.~J., {Kraus}, A., {Martinache}, F., {Law}, N., \& {Hillenbrand},
  L.~A. 2011, \apj, 726, 113

\bibitem[{{Ireland} {et~al.}(2008){Ireland}, {Kraus}, {Martinache}, {Lloyd}, \&
  {Tuthill}}]{2008ApJ...678..463I}
{Ireland}, M.~J., {Kraus}, A., {Martinache}, F., {Lloyd}, J.~P., \& {Tuthill},
  P.~G. 2008, \apj, 678, 463

\bibitem[{{Knight} {et~al.}(2012){Knight}, {Acton}, {Lightsey}, \&
  {Barto}}]{2012SPIE.8449E..0VK}
{Knight}, J.~S., {Acton}, D.~S., {Lightsey}, P., \& {Barto}, A. 2012, in
  Society of Photo-Optical Instrumentation Engineers (SPIE) Conference Series,
  Vol. 8449

\bibitem[{{Koekemoer} \& {Lindsay}(2005)}]{JWST.STScI.000647Koe}
{Koekemoer}, A.~M., \& {Lindsay}, K. 2005, in  (Baltimore: STScI)

\bibitem[{{Kraus} \& {Ireland}(2012)}]{2012ApJ...745....5K}
{Kraus}, A.~L., \& {Ireland}, M.~J. 2012, \apj, 745, 5

\bibitem[{{Lacour} {et~al.}(2011){Lacour}, {Tuthill}, {Amico}, {Ireland},
  {Ehrenreich}, {Huelamo}, \& {Lagrange}}]{2011AA...532A..72L}
{Lacour}, S., {Tuthill}, P., {Amico}, P., {et~al.} 2011, \aap, 532, A72

\bibitem[{{Lauer}(1999)}]{1999PASP..111.1434L}
{Lauer}, T.~R. 1999, \pasp, 111, 1434

\bibitem[{{Lightsey} {et~al.}(2004){Lightsey}, {Barto}, \&
  {Contreras}}]{2004SPIE.5487..825L}
{Lightsey}, P.~A., {Barto}, A.~A., \& {Contreras}, J. 2004, in Society of
  Photo-Optical Instrumentation Engineers (SPIE) Conference Series, Vol. 5487,
  Optical, Infrared, and Millimeter Space Telescopes, ed. J.~C. {Mather},
  825--832

\bibitem[{{Lloyd} {et~al.}(2006){Lloyd}, {Martinache}, {Ireland}, {Monnier},
  {Pravdo}, {Shaklan}, \& {Tuthill}}]{2006ApJ...650L.131L}
{Lloyd}, J.~P., {Martinache}, F., {Ireland}, M.~J., {et~al.} 2006, \apjl, 650,
  L131

\bibitem[{Macintosh {et~al.}(2014)Macintosh, Graham, Ingraham, Konopacky,
  Marois, Perrin, Poyneer, Bauman, Barman, Burrows, Cardwell, Chilcote,
  De~Rosa, Dillon, Doyon, Dunn, Erikson, Fitzgerald, Gavel, Goodsell, Hartung,
  Hibon, Kalas, Larkin, Maire, Marchis, Marley, McBride, Millar-Blanchaer,
  Morzinski, Norton, Oppenheimer, Palmer, Patience, Pueyo, Rantakyro, Sadakuni,
  Saddlemyer, Savransky, Serio, Soummer, Sivaramakrishnan, Song, Thomas,
  Wallace, Wiktorowicz, \& Wolff}]{Macintosh2014PNAS}
Macintosh, B., Graham, J.~R., Ingraham, P., {et~al.} 2014, Proceedings of the
  National Academy of Sciences

\bibitem[{{Macintosh} {et~al.}(2012){Macintosh}, {Anthony}, {Atwood},
  {Barriga}, {Bauman}, {Caputa}, {Chilcote}, {Dillon}, {Doyon}, {Dunn},
  {Gavel}, {Galvez}, {Goodsell}, {Graham}, {Hartung}, {Isaacs}, {Kerley},
  {Konopacky}, {Labrie}, {Larkin}, {Maire}, {Marois}, {Millar-Blanchaer},
  {Nunez}, {Oppenheimer}, {Palmer}, {Pazder}, {Perrin}, {Poyneer}, {Quirez},
  {Rantakyro}, {Reshtov}, {Saddlemyer}, {Sadakuni}, {Savransky},
  {Sivaramakrishnan}, {Smith}, {Soummer}, {Thomas}, {Wallace}, {Weiss}, \&
  {Wiktorowicz}}]{2012SPIE.8446E..1UM}
{Macintosh}, B.~A., {Anthony}, A., {Atwood}, J., {et~al.} 2012, in Society of
  Photo-Optical Instrumentation Engineers (SPIE) Conference Series, Vol. 8446

\bibitem[{{Makidon} {et~al.}(2008){Makidon}, {Sivaramakrishnan}, {Soummer},
  {Anderson}, \& {van der Marel}}]{2008SPIE.7010E..22M}
{Makidon}, R.~B., {Sivaramakrishnan}, A., {Soummer}, R., {Anderson}, J., \&
  {van der Marel}, R.~P. 2008, in Society of Photo-Optical Instrumentation
  Engineers (SPIE) Conference Series, Vol. 7010

\bibitem[{{Martinache}(2010)}]{2010ApJ...724..464M}
{Martinache}, F. 2010, \apj, 724, 464

\bibitem[{{Martinache}(2011)}]{2011SPIE.8151E..33M}
{Martinache}, F. 2011, in Society of Photo-Optical Instrumentation Engineers
  (SPIE) Conference Series, Vol. 8151

\bibitem[{{Martinache} \& {Guyon}(2009)}]{2009SPIE.7440E..0OM}
{Martinache}, F., \& {Guyon}, O. 2009, in Society of Photo-Optical
  Instrumentation Engineers (SPIE) Conference Series, Vol. 7440, Society of
  Photo-Optical Instrumentation Engineers (SPIE) Conference Series, 0

\bibitem[{{Martinache} {et~al.}(2007){Martinache}, {Lloyd}, {Ireland},
  {Yamada}, \& {Tuthill}}]{2007ApJ...661..496M}
{Martinache}, F., {Lloyd}, J.~P., {Ireland}, M.~J., {Yamada}, R.~S., \&
  {Tuthill}, P.~G. 2007, \apj, 661, 496

\bibitem[{{Martinache} {et~al.}(2009){Martinache}, {Rojas-Ayala}, {Ireland},
  {Lloyd}, \& {Tuthill}}]{2009ApJ...695.1183M}
{Martinache}, F., {Rojas-Ayala}, B., {Ireland}, M.~J., {Lloyd}, J.~P., \&
  {Tuthill}, P.~G. 2009, \apj, 695, 1183

\bibitem[{{Metchev} \& {Hillenbrand}(2004)}]{2004ApJ...617.1330M}
{Metchev}, S.~A., \& {Hillenbrand}, L.~A. 2004, \apj, 617, 1330

\bibitem[{{Monnier}(2003)}]{2003RPPh...66..789M}
{Monnier}, J.~D. 2003, Reports on Progress in Physics, 66, 789

\bibitem[{{Nielsen} {et~al.}(2013){Nielsen}, {Liu}, {Wahhaj}, {Biller},
  {Hayward}, {Close}, {Males}, {Skemer}, {Chun}, {Ftaclas}, {Alencar},
  {Artymowicz}, {Boss}, {Clarke}, {de Gouveia Dal Pino}, {Gregorio-Hetem},
  {Hartung}, {Ida}, {Kuchner}, {Lin}, {Reid}, {Shkolnik}, {Tecza}, {Thatte}, \&
  {Toomey}}]{2013ApJ...776....4N}
{Nielsen}, E.~L., {Liu}, M.~C., {Wahhaj}, Z., {et~al.} 2013, \apj, 776, 4

\bibitem[{{Oppenheimer} {et~al.}(2012){Oppenheimer}, {Beichman}, {Brenner},
  {Burruss}, {Cady}, {Crepp}, {Hillenbrand}, {Hinkley}, {Ligon}, {Lockhart},
  {Parry}, {Pueyo}, {Rice}, {Roberts}, {Roberts}, {Shao}, {Sivaramakrishnan},
  {Soummer}, {Vasisht}, {Vescelus}, {Wallace}, {Zhai}, \&
  {Zimmerman}}]{2012SPIE.8447E..20O}
{Oppenheimer}, B.~R., {Beichman}, C., {Brenner}, D., {et~al.} 2012, in Society
  of Photo-Optical Instrumentation Engineers (SPIE) Conference Series, Vol.
  8447

\bibitem[{{Perrin} {et~al.}(2012){Perrin}, {Soummer}, {Elliott}, {Lallo}, \&
  {Sivaramakrishnan}}]{2012SPIE.8442E..3DP}
{Perrin}, M.~D., {Soummer}, R., {Elliott}, E.~M., {Lallo}, M.~D., \&
  {Sivaramakrishnan}, A. 2012, in Society of Photo-Optical Instrumentation
  Engineers (SPIE) Conference Series, Vol. 8442

\bibitem[{{Pope} {et~al.}(2013){Pope}, {Martinache}, \&
  {Tuthill}}]{2013ApJ...767..110P}
{Pope}, B., {Martinache}, F., \& {Tuthill}, P. 2013, \apj, 767, 110

\bibitem[{{Readhead} {et~al.}(1988){Readhead}, {Nakajima}, {Pearson},
  {Neugebauer}, {Oke}, \& {Sargent}}]{1988AJ.....95.1278R}
{Readhead}, A.~C.~S., {Nakajima}, T.~S., {Pearson}, T.~J., {et~al.} 1988, \aj,
  95, 1278

\bibitem[{{Sabatke} {et~al.}(2005){Sabatke}, {Burge}, \&
  {Sabatke}}]{2005ApOpt..44.1360S}
{Sabatke}, E., {Burge}, J., \& {Sabatke}, D. 2005, \ao, 44, 1360

\bibitem[{{Sivaramakrishnan} {et~al.}(2009{\natexlab{a}}){Sivaramakrishnan},
  {Tuthill}, {Ireland}, {Lloyd}, {Martinache}, {Soummer}, {Makidon}, {Doyon},
  {Beaulieu}, \& {Beichman}}]{2009SPIE.7440E..30S}
{Sivaramakrishnan}, A., {Tuthill}, P.~G., {Ireland}, M.~J., {et~al.}
  2009{\natexlab{a}}, in Society of Photo-Optical Instrumentation Engineers
  (SPIE) Conference Series, Vol. 7440

\bibitem[{{Sivaramakrishnan} {et~al.}(2009{\natexlab{b}}){Sivaramakrishnan},
  {Tuthill}, {Martinache}, {Ireland}, {Lloyd}, {Perrin}, {Soummer}, {McKernan},
  \& {Ford}}]{2009astro2010T..40S}
{Sivaramakrishnan}, A., {Tuthill}, P., {Martinache}, F., {et~al.}
  2009{\natexlab{b}}, in ArXiv Astrophysics e-prints, Vol. 2010, astro2010: The
  Astronomy and Astrophysics Decadal Survey, 40

\bibitem[{{Sivaramakrishnan} {et~al.}(2010{\natexlab{a}}){Sivaramakrishnan},
  {Soummer}, {Oppenheimer}, {Carr}, {Mey}, {Brenner}, {Mandeville},
  {Zimmerman}, {Macintosh}, {Graham}, {Saddlemyer}, {Bauman}, {Carlotti},
  {Pueyo}, {Tuthill}, {Dorrer}, {Roberts}, \&
  {Greenbaum}}]{2010SPIE.7735E.266S}
{Sivaramakrishnan}, A., {Soummer}, R., {Oppenheimer}, B.~R., {et~al.}
  2010{\natexlab{a}}, in Society of Photo-Optical Instrumentation Engineers
  (SPIE) Conference Series, Vol. 7735

\bibitem[{{Sivaramakrishnan} {et~al.}(2010{\natexlab{b}}){Sivaramakrishnan},
  {Lafreni{\`e}re}, {Tuthill}, {Ireland}, {Lloyd}, {Martinache}, {Makidon},
  {Soummer}, {Doyon}, {Beaulieu}, {Parmentier}, \&
  {Beichman}}]{2010SPIE.7731E.126S}
{Sivaramakrishnan}, A., {Lafreni{\`e}re}, D., {Tuthill}, P.~G., {et~al.}
  2010{\natexlab{b}}, in Society of Photo-Optical Instrumentation Engineers
  (SPIE) Conference Series, Vol. 7731

\bibitem[{{Sivaramakrishnan} {et~al.}(2012){Sivaramakrishnan},
  {Lafreni{\`e}re}, {Ford}, {McKernan}, {Cheetham}, {Greenbaum}, {Tuthill},
  {Lloyd}, {Ireland}, {Doyon}, {Beaulieu}, {Martel}, {Koekemoer}, {Martinache},
  \& {Teuben}}]{2012SPIE.8442E..2SS}
{Sivaramakrishnan}, A., {Lafreni{\`e}re}, D., {Ford}, K.~E.~S., {et~al.} 2012,
  in Society of Photo-Optical Instrumentation Engineers (SPIE) Conference
  Series, Vol. 8442

\bibitem[{{Stephens} {et~al.}(2009){Stephens}, {Leggett}, {Cushing}, {Marley},
  {Saumon}, {Geballe}, {Golimowski}, {Fan}, \& {Noll}}]{2009ApJ...702..154S}
{Stephens}, D.~C., {Leggett}, S.~K., {Cushing}, M.~C., {et~al.} 2009, \apj,
  702, 154

\bibitem[{{Teague}(1982)}]{1982JOSA...72.1199T}
{Teague}, M.~R. 1982, Journal of the Optical Society of America (1917-1983),
  72, 1199

\bibitem[{{Thompson} {et~al.}(1986){Thompson}, {Moran}, \&
  {Swenson}}]{1986isra.book.....T}
{Thompson}, A.~R., {Moran}, J.~M., \& {Swenson}, G.~W. 1986, {Interferometry
  and synthesis in radio astronomy}

\bibitem[{{Troy} \& {Chanan}(2003)}]{2003ApOpt..42.3745T}
{Troy}, M., \& {Chanan}, G. 2003, \ao, 42, 3745

\bibitem[{{Tuthill} {et~al.}(2000){Tuthill}, {Monnier}, {Danchi}, {Wishnow}, \&
  {Haniff}}]{2000PASP..112..555T}
{Tuthill}, P.~G., {Monnier}, J.~D., {Danchi}, W.~C., {Wishnow}, E.~H., \&
  {Haniff}, C.~A. 2000, \pasp, 112, 555

\bibitem[{{Vigan} {et~al.}(2012){Vigan}, {Patience}, {Marois}, {Bonavita}, {De
  Rosa}, {Macintosh}, {Song}, {Doyon}, {Zuckerman}, {Lafreni{\`e}re}, \&
  {Barman}}]{2012AA...544A...9V}
{Vigan}, A., {Patience}, J., {Marois}, C., {et~al.} 2012, \aap, 544, A9

\bibitem[{{Wahhaj} {et~al.}(2013){Wahhaj}, {Liu}, {Nielsen}, {Biller},
  {Hayward}, {Close}, {Males}, {Skemer}, {Ftaclas}, {Chun}, {Thatte}, {Tecza},
  {Shkolnik}, {Kuchner}, {Reid}, {de Gouveia Dal Pino}, {Alencar},
  {Gregorio-Hetem}, {Boss}, {Lin}, \& {Toomey}}]{2013ApJ...773..179W}
{Wahhaj}, Z., {Liu}, M.~C., {Nielsen}, E.~L., {et~al.} 2013, \apj, 773, 179

\bibitem[{{Zimmerman} {et~al.}(2012){Zimmerman}, {Sivaramakrishnan}, {Bernat},
  {Oppenheimer}, {Hinkley}, {Lloyd}, {Tuthill}, {Brenner}, {Parry}, {Simon},
  {Krist}, \& {Pueyo}}]{2012SPIE.8445E..2GZ}
{Zimmerman}, N., {Sivaramakrishnan}, A., {Bernat}, D., {et~al.} 2012, in
  Society of Photo-Optical Instrumentation Engineers (SPIE) Conference Series,
  Vol. 8445

\bibitem[{{Zimmerman}(2011)}]{2011PhDT........54Z}
{Zimmerman}, N.~T. 2011, PhD thesis, Columbia University

\end{thebibliography}
%\input{ms_greenbaum112014.bbl}

\appendix

\section{A. The JWST NIRISS Non-Redundant Mask}

\begin{figure}[!htbp] \centering
\includegraphics[width=3.00truein]{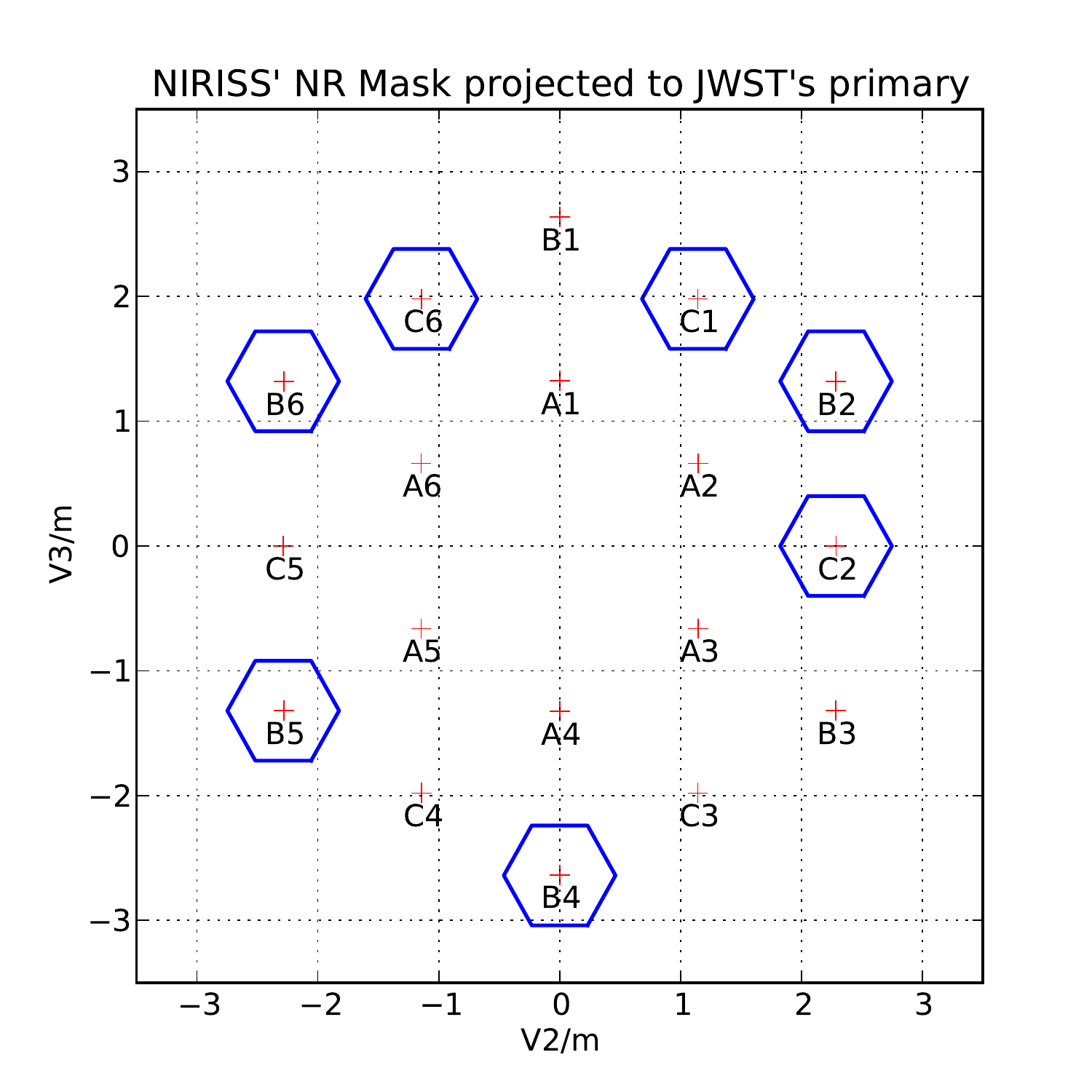}
\caption{\small \label{fig:maskcoords} The layout of the 7-hole NIRISS NRM
mask in JWST primary mirror coordinates (after \citet{2009SPIE.7440E..30S}).
The (V2,V3) axes are in the plane of the JWST pupil.
A viewer located at the secondary mirror and looking at the
reflective surface of the primary mirror would see this mask
projection and segment arrangement.
The centers of the 18 segments (designated A1-6, B1-6, and C1-6) are labelled.
The NRM's holes are nominally centered on the segment they expose.
The holes are 0.8~m flat-to-flat when projected back to the primary mirror.
The average segment flat-to-flat distance is approximately 1.32~m
if there were no inter-segment gaps.
The mask's holes are undersized so as to stay within the parent segment 
in the presence of a linear pupil misalignment 
of up to 3.8\% of the diameter of the pupil's circumscribing circle,
\viz\ 6603.5~mm at operating temperature
(Lightsey, private communication, Beaulieu, private communication).
All these numbers will need to be refined when the as-built pupil
distortion is measured on the ground and in flight.}
\end{figure}

NIRISS's non-redundant mask design exposes the central parts of seven of JWST's
segments in the outer ring of 12 segments.
\autoref{tab:NRMctrsPMV2V3} defines the nominal, as-designed, mask in primary
mirror space (shown in \autoref{fig:maskcoords}).  The mask's throughput is
approximately 15\% of the full aperture throughput (assuming spatially uniform primary
mirror reflectivity).  Thus the expected theoretical peak NRM PSF intensity
is $1/(0.15)^2 = 44$ times fainter than the corresponding the full aperture PSF.
In practise details of image centering, finite angular size of pixels,
filter bandpass, source spectrum, and the intra-pixel sensitivity
will cause slight deviations from this ratio.
NIRISS's F380M, F430, and 480M filters provide sufficiently fine sampling
on its $65~\mathrm{mas}$ pixel scale detector.
NIRISS is Nyquist sampled at $4~\micron$.
We have demonstrated reduced but still scientifically interesting capability for
NIRISS's NRM when used with the wide band F277W filter as well.

\capstartfalse
\begin{deluxetable}{crr}
\tabletypesize{\footnotesize}
\tablecolumns{3}
%\tablewidth{3.5truein}
\tablewidth{0pt}
\tablecaption{Nominal NRM hole centers in JWST primary mirror space\label{tab:NRMctrsPMV2V3}}
\tablehead{\colhead{Segment} & \colhead{V2/mm} & \colhead{V3/mm} }
\startdata
C1 &  1143 & 1980    \\
B2 &  2282 & 1317    \\
C2 &  2286 & 0       \\
B4 &  0    & -2635   \\
B5 &  2282 & -1317   \\
B6 &  2282 & 1317    \\
C6 &  1143 & 1980   
\enddata
\end{deluxetable}
\capstartfalse

\capstartfalse
\begin{deluxetable}{ccccccc}
\tabletypesize{\footnotesize}
\tablecolumns{7}
\tablewidth{0pt}
\tablecaption{ Estimated thermal background rates for \\
a ground-based 10~m telescope (left) and JWST NIRISS (right). \label{tab:Therm}}
\tablehead{
    \colhead{Wavelength} & \colhead{Bandwidth} & \colhead{Background}       & \colhead{\ \ \ \ \ \ } & \colhead{Wavelength} & \colhead{Bandwidth} & \colhead{Background}\\
    \colhead{\micron}    & \colhead{\%}        & \colhead{$e^-$s$^{-1}$pixel$^{-1}$}  & \colhead{}  & \colhead{\micron}     & \colhead{\%}        & \colhead{$e^-$s$^{-1}$pixel$^{-1}$} }
\startdata
 	 1.65    &  20\%  &  $5\times 10^{-4}$ & &  2.77   &  25\% &    0   \\
 	 2.20    &  20\%  &  $6\times 10^{-1}$ & &  3.80   &   5\% &    $9\times 10^{-2}$   \\
 	 3.50    &  20\%  &  $1\times 10^{3}$  & &  4.30   &   5\% &    $4\times 10^{-1}$   \\
 	 4.30    &  20\%  &  $1\times 10^{4}$  & &  4.80   &   8\% &    $5\times 10^{-1}$
\enddata
\tablecomments{ This rough estimate uses typical operating temperatures
    (273K for a ground-based telescope and 50K for JWST and NIRISS,
	and assumes that the thermal background for the ground-based telescope is entirely
	due to warm mirrors. On JWST NIRISS the entire opaque mask area
	will be the dominant source of thermal background.
	We use an emissivity of 0.1 for the warm Keck mirrors, and a system
	efficiency of 0.5 for both cases.}
\end{deluxetable}
\capstarttrue

\autoref{tab:Therm} shows a rough comparison of the estimated thermal
backgrounds assuming Keck-NIRC2's 9-hole mask and JWST NIRISS' 7-hole masks in many of the
filters that are used with their NRMs.  We assume that the ground-based mask itself is cooled,
so does not contribute to the thermal background.
Ground-based NRM is restricted to brighter adaptive optics guide stars,
is limited by thermal background longward of 3 \micron, but delivers better angular
resolution than NIRISS' NRM.
On the other hand, JWST NIRISS' NRM extends to 4.8~\micron\  with no appreciable
thermal background, so it should be able to observe much fainter targets than are
available to  instruments such as Keck-NIRC2 NRM.  We note that thermal background
limits Keck-NIRC2's $L'$ and $M_s$ filters to exposures of the order of 0.27 and 0.14 seconds respectively.

\end{document}